# Partition-Resolved Self-Diagnosing Quantum Hall Arrays using Star-Mesh Topologies

DaVonne Henry,[1] Dominick S. Scaletta,[2] Tamar M. Padwa,[3] Gretchen B. Ferguson,[4] Evelyn E. Hopkins,[4] Heather M. Hill,[1] Dean G. Jarrett,[1] David B. Newell,[1] and Albert F. Rigosi[1*]

[1]*Physical Measurement Laboratory, National Institute of Standards and Technology (NIST), Gaithersburg, MD 20899, United States*

[2]*Department of Physics, Mount San Jacinto College, Menifee, California 92584, United States*

[3]*Department of Physics, Georgetown University, Washington, DC 20057, United States*

[4]*Department of Physics, Auburn University, Auburn, AL 36832, United States*

ABSTRACT: The use of quantized Hall array resistance standards requires verifying that element quantization has not broken down without exhaustive per-element testing. Here, a Sequential Precision Quantization Reliability criterion is extended to arbitrarily partitioned, subarray-resolved topologies. By mapping resistance deviations to non-uniform observation windows, two-terminal boundary verification is enabled. Epitaxial graphene devices are measured to provide practical knowledge for intuiting acceptable levels of confidence derived from the SPQR criterion. Furthermore, these analyses demonstrate that star-mesh configurations suppress the probability of undetected breakdown by several orders of magnitude, comfortably approaching or surpassing a proposed 5σ metrological threshold.

[*] Email: afr1@nist.gov

## I. INTRODUCTION

Quantum Hall resistance standards have formed the basis of the metrological realization of the ohm for decades, and they provide stable plateaus at resistance values tied to multiples of the von Klitzing constant $R_{\mathrm{K}} = \frac{h}{e^2}$, with $h$ being the Planck constant and $e$ the elementary charge [1]. Recently, standards based on the $\nu = 2$ quantum Hall plateau in graphene ($R_H = h/2e^2 \approx 12906.4037\ \Omega$) have emerged as a preferred platform because it has less stringent requirements for quantization than standards based on GaAs heterostructures [2] – [7].

Despite the precision afforded by these devices, they are limited to a single value of resistance. For quantized values other than ≈12.9 kΩ, recent efforts have focused on developing arrays of interconnected graphene Hall elements, known as quantized Hall array resistance standards (QHARS) [8] – [12]. Superconducting NbTiN is used to link multiple Hall elements in series, parallel, or more sophisticated configurations that utilize Wye-Delta (Y-Δ) or star-mesh transformations [13] – [17]. The latter topologies are used to achieve broad scaling of resistance with a minimal increase in element count, largely advancing an objective to reduce the chain of traceability to $h$ and $e$.

Prior to international acceptance of QHARS as genuine standards, as was done with graphene-based single-element standards [7], one must be able to verify that quantization has not broken down across any element within the measurement path of an arbitrary QHARS device. Even with a reduction in element count, the element-by-element process of verification, which involves confirming the longitudinal resistance of the element ($\rho_{xx}$), becomes infeasible owing to the number of electrical leads required [7].

A recent introduction of a criterion for a QHARS device with a topological rotational symmetry demonstrated that a well-chosen set of measurements may be used to assure the lack of quantization breakdown to a high probability [13]. The framework presented herein expands on this initial understanding of the probability of artificiality by applying the same first principles to sequentially connected QHARS subarrays, and therefore, yields a quantity Φ that is labeled the Sequential Precision Quantization Reliability (SPQR) criterion. On top of the rotationally symmetric analogue of the SPQR criterion, these probabilistic confidence frameworks may also offer insights for statistically evaluating quantum technologies like topological qubit fidelities from input-output terminals.

## II. NUMERICAL AND EXPERIMENTAL METHODS

### A. Simulation Techniques

QHARS devices and proposed topologies were designed to output resistance values over six orders of magnitude with a preference for decadal values given their general compatibility with existing measurement infrastructure. Diagrams of the devices and proposed topologies used in this study are found in Figure 1 and include the following: (a) a three-partition ($n$ = 3) Y-Δ topology with two series arms and a single-element central grounded element (or shunt), (b) a four-partition ($n$ = 4) topology with three arms and a single grounded element, (c) a five-partition ($n$ = 5) two-stage decadic ladder with series segments and two single-element shunts, and (d) a five-partition ($n$ = 5) multi-tier star-mesh circuit using two 3-element lower-tier arms, two 31-element upper-tier arms, and an 8-element parallel shunt. The designs were modeled in LTspice to verify the expected quantized resistances [18 – 19].

### B. Experimental Information

The devices were fabricated with monolayer graphene grown epitaxially on 4H-SiC substrates. Using a polymer-assisted method, Si atoms are sublimated from SiC wafers in an inert environment of argon gas at 1900 °C [20]. The resulting graphene is patterned into devices by means of defining a Pd/Au etch mask in the shape of the desired devices and defining NbTiN superconducting interconnects using photolithography [16 - 17], [20]. The carrier density of the fabricated devices was tuned after functionalization with $Cr(CO)_3$ [21, 22], and electrical characterization was performed using a Keithley 2612A Source Measure Unit in a cryogenic system at 1.6 K (see Acknowledgments).

To define experimental observation windows, a dual source bridge (DSB), or Wheatstone-style bridge, is used at and above MΩ levels, and it can achieve noise levels slightly under 100 nΩ/Ω [23] – [25]. To precisely determine the QHARS device values using the DSB, 1 MΩ and 10 MΩ standard resistors (Fluke 742A) were used, whose exact values were determined at around the time of the QHARS device measurements.

## III. SEQUENTIAL PRECISION QUANTIZATION RELIABILITY CRITERION FOR SUBARRAY-RESOLVED PARTITIONS

Consider an interconnected QHARS device where the individual quantum Hall elements are organized into characterizable partitions. Each partition has one or more quantum Hall element connected in series or parallel. The SPQR criterion ($\Phi$) indicates a confidence level (percent probability) that a QHARS has not experienced quantization breakdown. Previously, a similar criterion was developed for QHARS with rotationally symmetric topologies about a central node (cyclic group $C_n$) [13]. Beginning in a similar vein to that work, an internal elemental defect $\delta_i$ is incorporated into the resistance of the $i$-th quantum Hall element such that the Hall resistance of element $i$ is $R_i = R_H + \delta_i$. Because of thermodynamic irreversibility with respect to dissipation, $\delta_i \geq 0$. A microscopic breakdown parameter $\mu_i$ represents the longitudinal resistivity exhibited by an element that has experienced a phase transition out of the quantum Hall effect and has been reported as about 1 Ω [2], [26] – [28]. From the elemental scale breakdown, a partition-level breakdown parameter $\mu_j$ (or $\mu_n$ when uniform across all $n$ partitions) is defined as the internal resistance shift of a partition due to a single element failing, while the circuit-level star-mesh amplification between partitions and two-terminal measurements is captured directly in the sensitivity matrix $\boldsymbol{S}$ (unlike rotationally symmetric topologies [13], where both were bundled into $\mu_n$; see Supplemental Material for details [29]). When a partition contains multiple internal segments, the partition-level breakdown parameter can have different values based on where the breakdown occurs. For a conservative approach to calculating Φ, a breakdown parameter is calculated for a defect occurring in each segment, and the smallest breakdown value is selected for the SPQR calculation since that value will result in the smallest SPQR value.

A defect scalar representing the smallest partitions of a device whose defective nature could be identified is included in the total resistance $R_j = R_{\text{nominal}} + \delta_j$ (where $\delta_j$ is a quantization error or defect resistance scalar and $j \in \{1, \dots, n\}$). The subscript $n$ represents the number of unique device partitions, measurements across which are essential for defining a probability that all measurements, if in strong agreement, are a result of artificial (not purely quantized) precision.

The deviation in electrical resistance from the quantized value of the $k$-th measurement configuration may be expressed as a combination of the $\delta_j$ defects in a sensitivity matrix $\boldsymbol{S}$: $\Delta R_k = \sum_{j=1}^{n} S_{k,j}\, \delta_j$ (for $k = 1, \dots, N$). If $N$ independent measurements are performed in a QHARS device topology in which $n$ unique partitions reside ($N \geq n$), then the SPQR criterion approaches 1 with increasing $N$, indicating better QHARS quantization confidence.

As in [13], a Laplace distribution is assumed for the defect scalars which are used to define a joint probability density function $f(\vec{\delta})$. The result is similar to the case of rotationally symmetric topologies [13]:

$$f(\vec{\delta}) = \prod_{j=1}^{n} \frac{1}{2\mu_j} \exp\left(-\frac{|\delta_j|}{\mu_j}\right) = \frac{1}{2^n \prod_{j=1}^{n} \mu_j} \exp\left(-\sum_{j=1}^{n} \frac{|\delta_j|}{\mu_j}\right) \tag{1}$$

The probability of artificiality $P(A)$ is defined as the likelihood that an internal defect state exists and that it falls within an observation window $\epsilon$ (the standard deviation of repeated measurements of configuration $k$). By constructing an integral over a space of observables [13]:

$$P(A) = \frac{1}{2^n \sqrt{\det(\mathbf{S}^T\mathbf{S})} \prod_{j=1}^{n} \mu_j} \int_{\mathcal{P}} \exp\left(-\sum_{j=1}^{n} \frac{\left|\left[(\mathbf{S}^T\mathbf{S})^{-1}\mathbf{S}^T \Delta\vec{R}\right]_j\right|}{\mu_j}\right) d^n(\Delta R) \tag{2}$$

When all partitions share a uniform two-terminal breakdown scale $\mu_j = \mu_n$, the prefactor reduces to $\left[\sqrt{\det(\mathbf{S}^T\mathbf{S})}\,(2\mu_n)^n\right]^{-1}$ with exponent $-\frac{1}{\mu_n}||(\mathbf{S}^T\mathbf{S})^{-1}\mathbf{S}^T\Delta\vec{R}||_1$, much like what is seen in the rotationally symmetric case [13]. A key physical difference in Eq. 2 is in how $n$ is defined, and that for the purposes herein, $n$ is the unique number of partitions. How Eq. 2 is evaluated depends on the topology being adopted, as well as which measurements are being collected. Since these decisions may lead to ambiguity in how the SPQR criterion can be evaluated, two example approaches, one based on the subarrays and the other based on Y-Δ configurations, will be compared to elaborate benefits seen with the stronger of the two approaches for calculating the SPQR criteria.

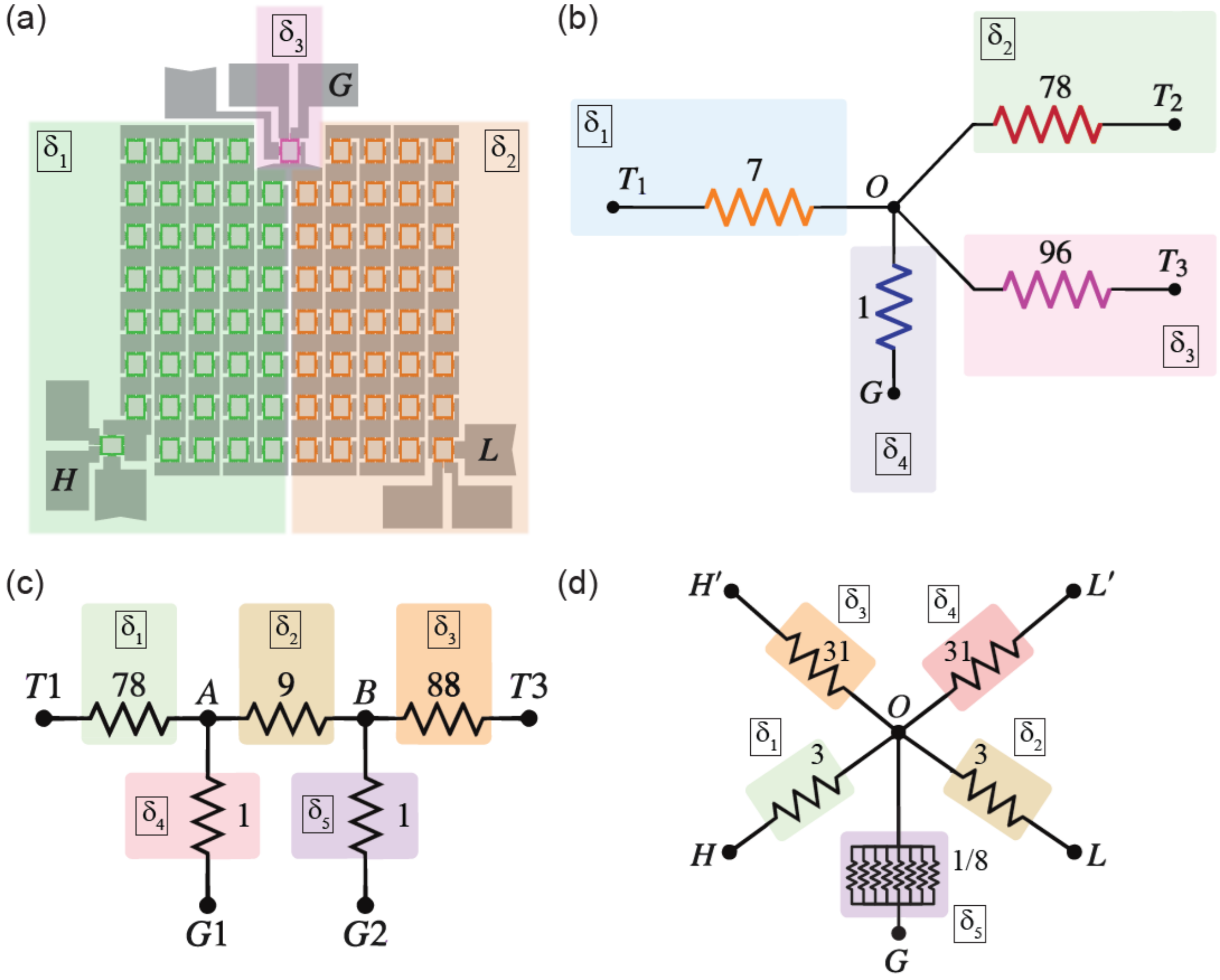

FIG. 1. (Color online) Subarray-resolved QHARS device topologies and defect partitions. (a) Fabricated three-partition ($n$ = 3) symmetric Y-Δ QHARS topology consisting of two series arms with 39 elements per arm and a single-element central shunt. (b) Proposed four-partition ($n$ = 4) array comprising three arms and a single-element central shunt, yielding near-decadic outputs across five orders of magnitude. (c) Proposed five-partition ($n$ = 5) two-stage decadic ladder with series segments and two single-element shunts, spanning six orders of magnitude. (d) An alternative proposed five-partition ($n$ = 5) multi-tier star-mesh circuit resolving two 3-element lower-tier arms, two 31-element upper-tier arms, and an 8-element parallel shunt.

### A. SPQR Criterion via Subarray Measurements

For subarray-resolved SPQR criteria, $\delta_n$ are defects present in each of the smallest measurable partitions. This first approach proposes measurements that involve two defect vectors, and there are three unique two-terminal measurements, per Fig. 1 (a): Across the regions represented by (1) $\delta_1 + \delta_2$, (2) $\delta_1 + \delta_3$, and (3) $\delta_2 + \delta_3$. Unlike a rotationally symmetric topology [13], each measurement is expected to yield different quantized values, compelling one to reevaluate the SPQR criterion's integration process.

To account for different nominal values, and thus potentially different noise floors, the hypervolume of the integration bounds of Eq. 2 must reflect a non-uniform observation window $\epsilon$ across the measurements. Examples of multi-partitioned arrays are proposed in Fig. 1 (b)-(d). If one continues to assume that $\epsilon_i \ll \mu_n$ [13], then the geometric volume of the acceptance region in measurement space transforms from an $n$-dimensional hypercube into an orthotope (a hyperrectangle). The volume of the observation space thus becomes the product of these three distinct observation windows, leading to the following adjustment of the SPQR criterion:

$$\Phi \approx 1 - \frac{1}{\sqrt{\det(S^T S)}}\left(\frac{\epsilon_1 \epsilon_2 \epsilon_3}{{\mu_n}^3}\right) = 1 - \frac{1}{2}\frac{\epsilon_1 \epsilon_2 \epsilon_3}{{\mu_n}^3} \tag{3}$$

See Supplemental Material for more details on formulating Eq. 3 [29].

### B. SPQR Criterion via Y-Δ Measurements

As an alternative approach, one can include a measurement of the Y-Δ configuration using the same defect vectors $\delta_n$. For the example in Fig. 1 (a), the configuration could require two-terminal measurements to be made across the entire array while grounding the single element shunt (characterized by $\mu_s$), yielding a different sensitivity matrix $\boldsymbol{S}$ that incorporates the circuit-level star-mesh amplification ($|\det(\boldsymbol{S}')| = 1521$). Allowing distinct partition breakdown parameters $\mu_a$ and $\mu_s$ for the series arms and central shunt leads to a similar small-$\epsilon$ derivation of the SPQR criterion:

$$\Phi \approx 1 - \frac{\epsilon_\Delta \epsilon_{78} \epsilon_{40}}{1521\, \mu_a^2 \mu_s} \tag{4}$$

More information on the specific steps to obtain Eq. 4 can be found in the Supplemental Material [29] (where $\mu_a = 1\ \Omega$ and $\mu_s = 1\ \Omega$ are the internal partition breakdown parameters of the series arms and single-element shunt, respectively, while the Y-Δ shunt amplification resides in $\det(\boldsymbol{S}')$). Since $\epsilon$ is expected to take on ideally similar orders of magnitude, this evaluation suggests that including a Y-Δ measurement that is in strong agreement with its nominal value (i.e., has a small observation window $\epsilon$) is likely to provide stronger confidence that a quantization breakdown did not occur throughout an array. For the particular example in Fig. 1 (a) (and Eq. 4), $\epsilon_\Delta$, $\epsilon_{78}$, and $\epsilon_{40}$ represent the observation windows for the Y-Δ, full series array, and near-half-series arrays, respectively. The ratio between Eq. 3 and 4 is thus hypothesized to favor Y-Δ cases by a few orders of magnitude in $P(A)$.

### C. Exact SPQR Integration Techniques

For observations of $\Delta\vec{R}$ that do not satisfy the condition $\epsilon \ll \mu_n$, a more exact approach must be taken. For instance, if $n = 3$, then one can expand Eq. 1 and Eq. 2 to obtain:

$$P(A) = \frac{1}{8\sqrt{\det(\mathbf{S}^T\mathbf{S})}\ \mu_1\mu_2\mu_3}\int_{-\epsilon_1}^{\epsilon_1}\int_{-\epsilon_2}^{\epsilon_2}\int_{-\epsilon_3}^{\epsilon_3}\exp\left[-\frac{|\Delta R_1 + \Delta R_2 - \Delta R_3|}{2\mu_1} - \frac{|\Delta R_1 - \Delta R_2 + \Delta R_3|}{2\mu_2} - \frac{|-\Delta R_1 + \Delta R_2 + \Delta R_3|}{2\mu_3}\right] d\Delta R_3\, d\Delta R_2\, d\Delta R_1 \quad (5)$$

If all breakdown parameters are identical ($\mu_j = \mu_n$), one can evaluate each inner integral until a final piecewise expression for unequal $\epsilon_j$ can be formulated (see Supplemental Material [29]). Additionally, for the convenient case where all observations windows are equal, Eq. 5 may be reduced to:

$$P(A) = 1 - \frac{1 + 3\left(\frac{\epsilon}{\mu_n}\right)}{2} e^{-\left(\frac{\epsilon}{\mu_n}\right)} - e^{-\frac{3}{2}\left(\frac{\epsilon}{\mu_n}\right)} + \frac{3}{2} e^{-2\left(\frac{\epsilon}{\mu_n}\right)} - e^{-\frac{5}{2}\left(\frac{\epsilon}{\mu_n}\right)} \quad (6)$$

The expected limiting behavior emerges when $\left(\frac{\epsilon}{\mu_n}\right) \ll 1$, whose series expansion returns:

$$P(A) = \frac{1}{\sqrt{\det(\boldsymbol{S}^T\boldsymbol{S})}}\left(\frac{\epsilon}{\mu_n}\right)^3 + O\left[\left(\frac{\epsilon}{\mu_n}\right)^4\right]$$

## IV. SPQR CRITERION FOR GRAPHENE-BASED DEVICES

To evaluate how subarray-resolved SPQR criteria behave with partition multiplicity ($n$) and various measurement equipment limits, two EG-based QHARS devices were fabricated and measured (Fig. 1 (a) and Fig. 2). Three EG-based QHARS designs were subsequently proposed and analyzed under ideal experimental noise floors (see Fig. 1 (b)-(d)).

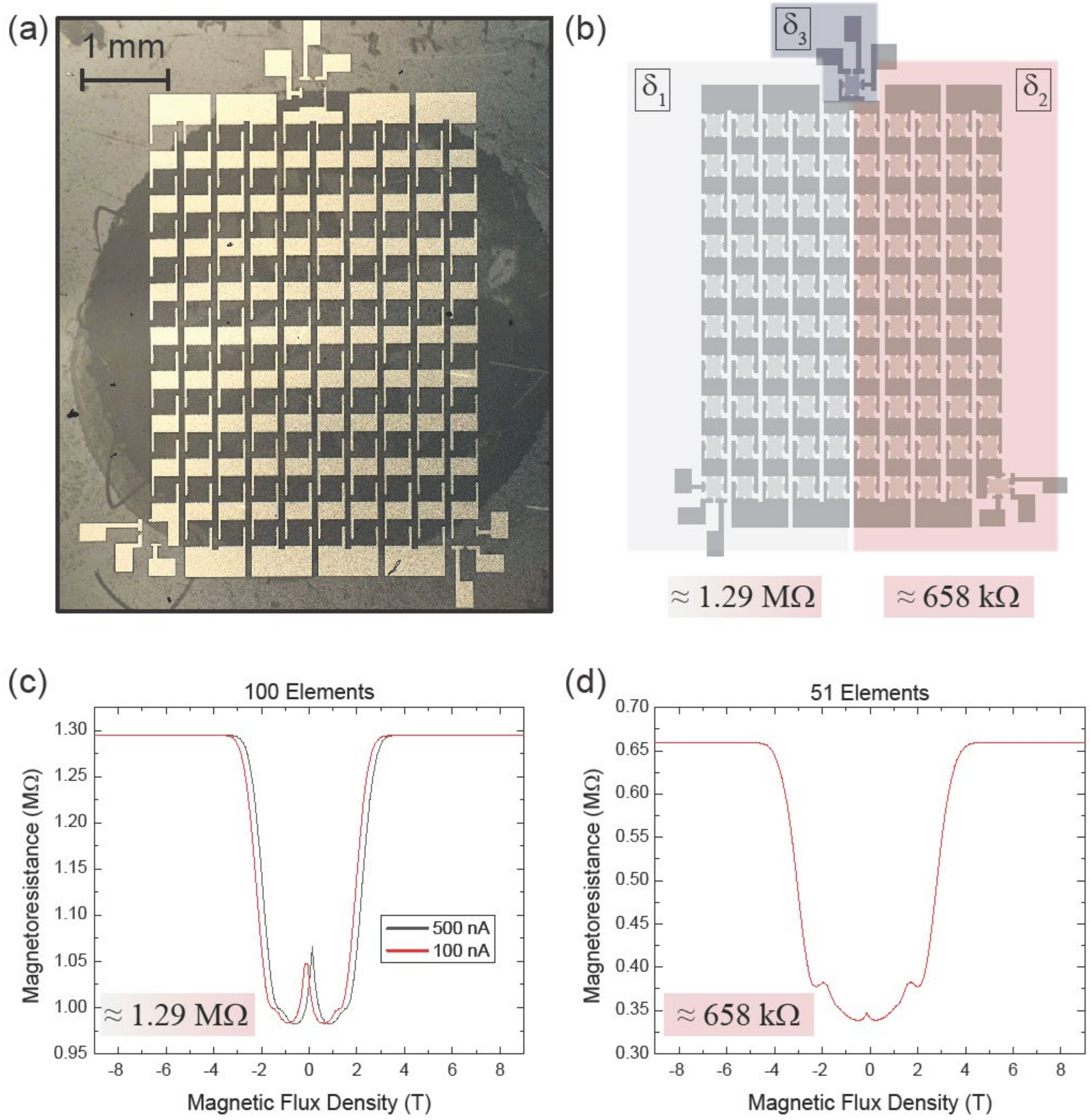

FIG. 2. (Color online) One of two nanofabricated QHARS devices, originally designed to pursue Y-Δ configured quantized resistance measurements. (a) Optical image of finished device prior to wire bonding. (b) Layout design with defect vector assignments as required by the SPQR criterion analysis. (c) Magnetoresistance at two applied currents across the 100-element series subarray, accounting for basic leakage through measurement setup. (d) Magnetoresistance at 500 nA for a 51-element in series subarray, justifying the need for more precise measurements.

### A. SPQR Criterion for Suboptimal Measurements

To establish a baseline illustrating the necessity of metrological-grade resistance bridges when applying the SPQR framework, the 101-element QHARS device shown in Fig. 2 (two 50-element series arms and a single-element central shunt), corresponding to $n = 3$ partitions, was first characterized using a source-measure unit (Keithley 2612A, see Acknowledgments). Three two-terminal subarray configurations were measured across applied currents from 5 μA to 20 μA (Fig. 3): (1) the full 100-element series path across both arms and (2) the two symmetric 51-element series paths pairing each arm with the single element shunt.

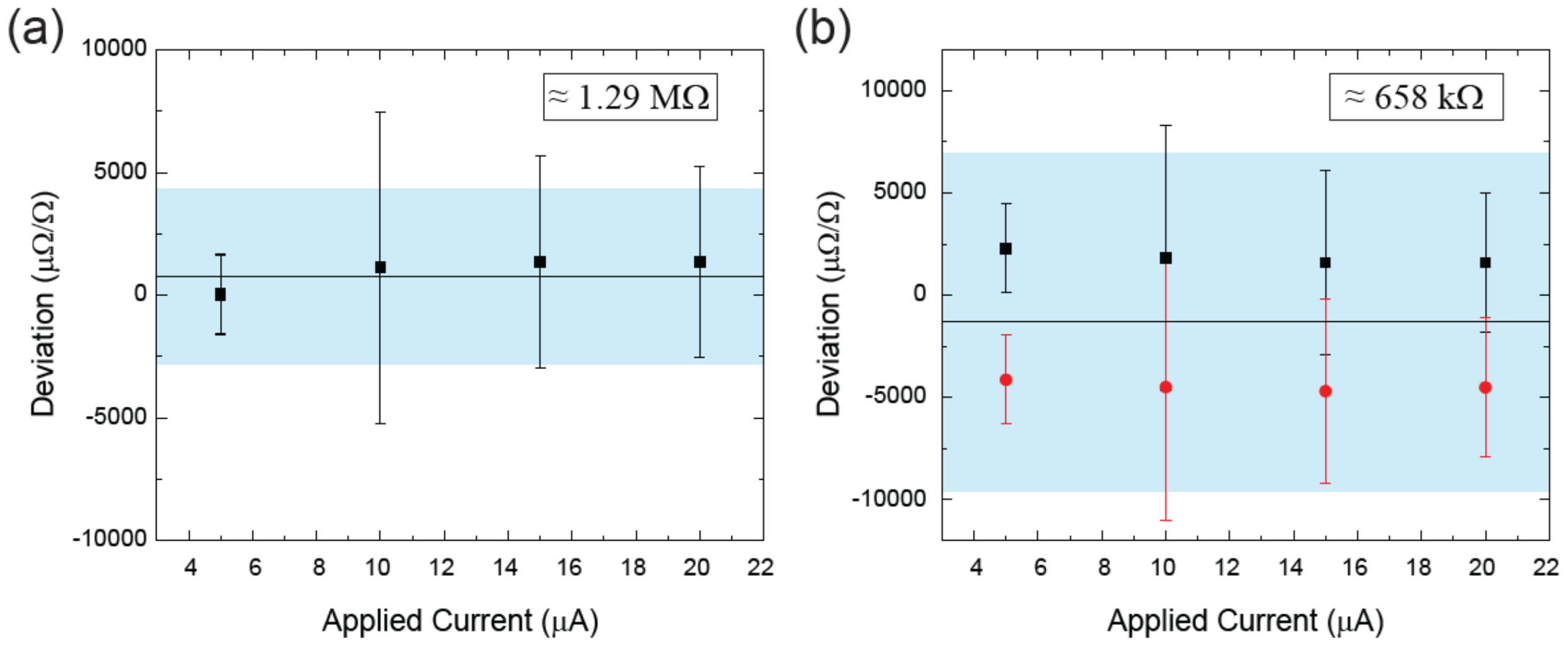

FIG. 3. (Color online) Two-terminal resistance measurements of the 100-element QHARS device subarray (from Fig. 2) as a function of applied current. (a) Relative resistance deviation from nominal (about 1.29 MΩ) across the full 100-element series array, yielding a measurement standard deviation of 3564.122 μΩ/Ω (shaded blue region, corresponding to $\epsilon_{100} \approx 4600$ Ω, and includes the individual error bar contributions from each point). (b) Relative resistance deviation from nominal (about 658 kΩ) across the two 51-element arm-plus-shunt subarrays (black squares and red circles), yielding a combined measurement standard deviation of 8290.45817 μΩ/Ω (shaded blue region, corresponding to about $\epsilon_{51} \approx 5457$ Ω, and includes the individual error bar contributions from each point).

Converting the empirical relative standard deviations in Fig. 3 into absolute observation windows $\epsilon = R_{\text{nominal}} \times \sigma_{\text{rel}}$ yields values that exceed the elemental breakdown scale ($\mu_n = 1\ \Omega$) by more than three orders of magnitude. The small-$\epsilon$ orthotope approximation of Eq. 3 is inapplicable and the exact integration of Eqs. 5 and 6 must be employed. In the limit where $\epsilon_{100}, \epsilon_{51} \gg \mu_n$, all exponential decay terms $\exp(-\epsilon_k/\mu_n)$ vanish identically, causing the triple integral in Eq. 5 to saturate at unity: $P(A) \rightarrow 1$, corresponding to an SPQR confidence level of 0 %. Physically, because a single-element breakdown shift ($\mu_n \approx 1\ \Omega$) is completely buried within the kΩ voltmeter noise floor, suboptimal two-terminal measurements cannot statistically exclude internal quantization breakdown.

### B. Comparing SPQR Criteria for Star-Mesh Configurations

To demonstrate how precision bridge measurements and star-mesh (Y-Δ) configurations dramatically strengthen the SPQR criterion, the 79-element QHARS device of Fig. 1 (a) (two 39-element series arms ($\delta_1, \delta_2$) and a single-element central shunt ($\delta_3$)) was evaluated using a dual source bridge (DSB), as shown in Fig. 4. The relative standard deviations of the time-series measurements for the 40-element, 78-element, and 79-element Y-Δ configuration are about 30.5 $\mu\Omega/\Omega$, 1.257 $\mu\Omega/\Omega$, and 10.7 $\mu\Omega/\Omega$, respectively, yielding $\epsilon_{40} = 15.75\ \Omega$, $\epsilon_{78} = 1.27\ \Omega$, and $\epsilon_{\Delta} = 220.82\ \Omega$.

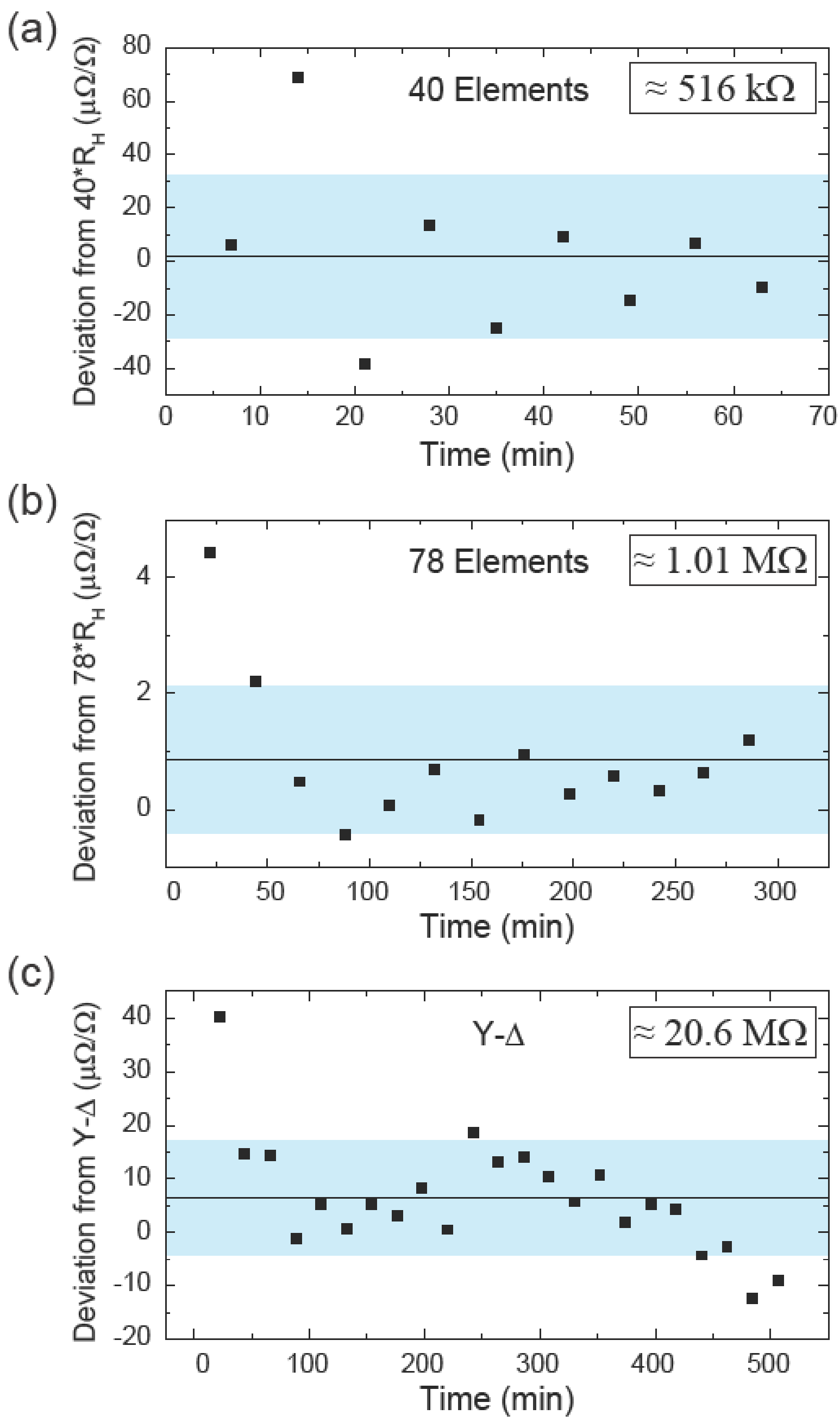

FIG. 4. (Color online) Dual source bridge time-series resistance measurements for the 79-element Y-Δ QHARS device of Fig. 1 (a). Relative resistance deviations from respective nominal values are shown for (a) a 40-element arm-plus-shunt subarray, exhibiting a standard deviation of about 30.5 μΩ/Ω (shaded blue region, corresponding to $\epsilon_{40}$). (b) the 78-element full series path, exhibiting a standard deviation of about 1.257 μΩ/Ω (shaded blue region, corresponding to $\epsilon_{78}$). (c) the Y-Δ output (about 20.6 MΩ) with the central single-element shunt

grounded, exhibiting a standard deviation of 10.7 μΩ/Ω (shaded blue region, corresponding to $\epsilon_\Delta$).

First, compute the subarray-only SPQR criterion using the DSB measurements of Figs. 4 (a) and 4 (b) alone (applying $\epsilon_{40} = 15.75\ \Omega$ twice alongside $\epsilon_{78} = 1.27\ \Omega$). Because $\epsilon_{40}$ and $\epsilon_{78}$ exceed the unit series breakdown scale $\mu_n = 1\ \Omega$, the exact integration of Eq. 5 yields $P(A)_{\text{sub}} = 0.5408$, ($\Phi_{\text{sub}} = 45.92$ %). The second SPQR criterion incorporates the Y-Δ measurement in Fig. 4 (c) ($\epsilon_\Delta = 220.82\ \Omega$) together with Figs. 4 (a) and 4 (b) using the minimal full-rank $3 \times 3$ sensitivity matrix $\mathbf{S}'$ ($|\det(\mathbf{S}')| = 1521$). The exact integration is shown in the Supplemental Material and yields $\Phi_{\text{Y-}\Delta} = 0.927$. Furthermore, at ideal bridge noise floors ($100\ \text{n}\Omega/\Omega$ DSB for the Y-Δ path and a CCC for $\epsilon_{40}$, where $\epsilon_k \ll \mu_n$ holds across all three measurements), Eq. 4 predicts $P(A)_{\text{Y-}\Delta} = 7.05 \times 10^{-8}$ ($\Phi_{\text{Y-}\Delta} = 0.99999993$, surpassing the proposed metrological $5\sigma$ threshold).

### C. SPQR Criteria for Ideal Multi-Decadal QHARS Topologies

To assess how subarray-resolved SPQR criteria scale across broader resistance ranges and higher partition multiplicities ($n = 4$ and 5) under ideal equipment noise floors, three proposed topologies were theoretically evaluated (see Figs. 1 (b)-(d)). These evaluations, largely detailed in the Supplemental Material [29], demonstrate how multi-stage and multi-tier star-mesh topologies can simultaneously expand the accessible decadic resistance range and strengthen self-diagnostic fault detection across bridge regimes (only DSB bridges were used for experimental data in this work).

In the first five-partition architecture (Fig. 1 (c)), even though a 2.85× improvement is predicted for star-mesh over subarray-only measurements, all SPQR criteria are predicted to exceed the 5σ proposed confidence level when measurements are taken at ideal noise floors [13, 30]. This behavior may readily be attributed to the number of partitions and is analogous to the behavior seen with higher orders of rotational symmetry in prior work [13]. In the second five-partition topology (Fig. 1 (d)), similar heightened levels of SPQR criterion are attainable merely by measuring a high number of partitions. All of the examples indicate that a sufficiently high confidence may be engineered in ways that are less restrictive than rotationally symmetric topologies, but careful attention is required to ensure such multi-decadal QHARS devices (or more generally, QHARS devices with non-uniform observation windows) may be compatible with multiple types of bridges and measurement apparatuses.

## V. CONCLUSIONS

In summary, the Sequential Precision Quantization Reliability (SPQR) framework has been generalized from rotationally symmetric cyclic topologies to arbitrarily partitioned, subarray-resolved quantum Hall resistance networks, primarily to include non-uniform observation windows and exact analytical integration methods for regimes where measurement uncertainty exceeds the elemental breakdown scale. Experimental comparisons on nanofabricated epitaxial graphene QHARS devices demonstrate that while coarse source-measure unit characterization saturates at the uninformative limit ($\Phi \rightarrow 0.0$ %), precision dual source bridge measurements incorporating Y-Δ (star-mesh) configurations amplify internal shunt defect sensitivities such that the probability of hidden quantization breakdown is suppressed by several orders of magnitude compared to series subarray measurements alone.

Theoretical evaluations of four- and five-partition star-mesh QHARS topologies confirm that single-chip devices can simultaneously deliver decadic resistance outputs spanning up to six orders of magnitude while exceeding the proposed 5σ metrological reliability threshold. These efforts may also have their topological fault-amplification principles adapted to low-pin-count diagnostic verification in emerging quantum information and nanoelectronic circuits.

## ACKNOWLEDGMENTS

Unless otherwise noted, NIST work was funded solely by the United States Government. The authors declare no competing interest.

The authors also wish to express their appreciation to P. Barbara at Georgetown University and Y. Yang at NIST for fruitful discussions. D. Henry would like to thank the National Research Council's Research Associateship Program for the opportunity.

# Supplemental Material

# Partition-Resolved Self-Diagnosing Quantum Hall Arrays using Star-Mesh Topologies

DaVonne Henry,[1] Dominick S. Scaletta,[2] Tamar M. Padwa,[3] Gretchen B. Ferguson,[4] Evelyn E. Hopkins,[4] Heather M. Hill,[1] Dean G. Jarrett,[1] David B. Newell,[1] and Albert F. Rigosi[1†]

[1]*Physical Measurement Laboratory, National Institute of Standards and Technology (NIST), Gaithersburg, MD 20899, United States*

[2]*Department of Physics, Mount San Jacinto College, Menifee, California 92584, United States*

[3]*Department of Physics, Georgetown University, Washington, DC 20057, United States*

[4]*Department of Physics, Auburn University, Auburn, AL 36832, United States*

Table of Contents:

---

**Table of Definitions and High-Level Description of the SPQR Criterion**

The Sequential Precision Quantization Reliability (SPQR) criterion is a percentage confidence level that a given quantized Hall array resistance standard (QHARS) topology exists such that none of its elements have experienced a quantum phase transition away from the quantized Hall effect. The definitions of significant variables include:

[1] $\Phi$ – Sequential Precision Quantization Reliability (SPQR) criterion, acting as a statistical threshold of acceptance or confidence level that all QHARS device elements have not exited the quantum Hall state (abstractly identical to the Symmetrical Precision Quantization Reliability criterion).

[2] $n$ – Distinct partitions for a device's topology.

[3] $\delta_n$ – The macroscopic quantization error, or defect resistance scalar, for a single subarray. The full vector is $\vec{\delta}$.

[4] $\mu_n$ – Macroscopic subarray breakdown parameter.

[5] $\epsilon$ – The observation window, corresponding to the standard deviation of the $N$ measurements.

[6] $N$ – Number of distinct measurements performed in an $n$-partitioned topology.

[7] $\Delta R_N$ – The deviation in resistance of the $N$-th measurement configuration.

[8] $S$ – Sensitivity matrix containing elements $S_{N,j}$, which expresses the deviation in measurement as a combination of defects.

[9] $P(A)$ – The probability of artificiality, defined as the probability that an internal defect state or lack of quantization exists and lands inside the observation window $\epsilon$.

## 1. Additional Experimental and Theoretical Details

Measurements were performed for two arrays in the main text using a dual source bridge (DSB). Though the Keithley 2612A was used for preliminary measurements, it did not contribute to high precision results due to limitations on the equipment (leading to mΩ/Ω errors). The devices under test were measured against a calibrated reference standard (Fluke 742A). The bridge was operated in a 1:1 range (or 2.06:1 for the Wye-Delta) configuration.

As with the messaging in the work on rotationally symmetric topologies, on the issue of selecting the Laplace distribution, which also matches the ISO/IEC Guide 98-3:2008/Suppl 1:2008, we wish to explain that the probability density function $f(\delta)$ in this framework is not used to model the empirical variance of our measurement data (which is captured separately by the observation window $\epsilon$). It instead models the unobservable, theoretical state space of hidden internal quantization defects. We are thus constrained to physical priors, namely, that we know the expected characteristic scale of a failed element ($\mu_i \approx 1\ \Omega$), but we do not possess a statistically well-behaved variance ($\sigma^2$) for these rare macroscopic faults. In the context of quantum electrical metrology, adopting a Gaussian distribution would aggressively suppress the mathematical likelihood of severe, singular macroscopic faults due to its $e^{-x^2}$ tail decay, whereas the Laplace distribution accommodates heavier tails ($e^{-|x|}$), providing a conservative upper bound for $P(A)$.

To further explain the joint probability density function, in statistical mechanics, a state space must be defined containing every possible configuration of a system, and then a probability must be assigned to each state. Therefore, $f(\delta)$ maps this defect space of the entire QHARS device. It answers the following question: "What is the exact probability that the whole design exists in a specific configuration wherein macroscopic errors exist?" Before looking at the whole device, one must first focus on just one partition. This partition (or arm) is assigned a Laplace distribution, whose shape indicates that when one has many measurements in strong agreement, the most likely physical scenario is that the arm is fully quantized ($\delta_j = 0$). If a defect does exist, it could manifest as a positive or negative macroscopic shift, with the probability of that shift occurring decaying exponentially as the shift gets larger. The width of the distribution for each partition is determined by its internal partition breakdown parameter $\mu_j$ (or $\mu_n$ when uniform across all partitions), while circuit-level star-mesh amplification is captured when mapping $\vec{\delta}$ through the sensitivity matrix $\boldsymbol{S}$.

The next step is to scale up to the entire device. The array has $n$ independent partitions. Because the failures are independent, to find the joint probability of the entire device's state, one multiplies the individual probabilities of all $n$ partitions together:

$$f(\vec{\delta}) = \prod_{j=1}^{n} \frac{1}{2\mu_j} \exp\left(-\frac{|\delta_j|}{\mu_j}\right) = \frac{1}{2^n \prod_{j=1}^{n} \mu_j} \exp\left(-\sum_{j=1}^{n} \frac{|\delta_j|}{\mu_j}\right) \tag{S1}$$

Equation S1 looks similar to a Boltzmann distribution ($e^{-E/kT}$), and so one can think of the sum term, $\sum|\delta_j|$, as the total defect energy of the entire QHARS device. The equation thus states that the probability of the entire QHARS device existing in one specific defective state drops exponentially based on the sum total of all the macroscopic errors present across the structure. Overall, this approach mathematically models the entire $n$-dimensional space of every possible way the device can fail. The SPQR criterion then proceeds to calculate what fraction of that failure volume can hide inside the noise limit ($\epsilon$) (since anything outside that noise limit can presumably be measured and thus potentially accounted for), and the larger that fraction, the lower the confidence that an element in the QHARS device has not broken down.

## 2. Discussion on $\vec{\delta}$

This section will focus on demonstrating how one translates the defect vector of a single element to the corresponding macroscopic version. It has been presented in the Supplemental Material of a previous work and calls upon a prior topology [1]. This procedure maps the complex physical failure modes of the device into a single aggregated degree of freedom per branch. Note that the analysis in this section is not required for the general use of the SPQR criterion; it is provided for purposes of edification.

Since translations between elemental and macroscopic quantities is dependent on the exact layout that is used, this example will focus on the 1 GΩ design presented in Ref. [1] (where each of the four branches may be labeled as North, East, South, and West). Recall that $\delta_i$ captures the small, localized deviations or non-idealities within individual components of the measurement network.

To begin, the high-dimensional internal defect space must be projected onto a reduced $n$-dimensional manifold (these $n$ dimensions may either represent the order of rotational symmetry as in [1] or the number of unique QHARS device partitions). First define a strictly additive macroscopic segment defect, $\delta_n$, for each branch. From the terminal to a central node (and/or central ground), the effective resistance of a branch (such as the North quadrant) is modeled as $R_{\text{North}} = R_{\text{North, nominal}} + \delta_{\text{North}}$. Here, $\delta_{\text{North}}$ is an encapsulation of every microscopic degree of freedom within that branch. Recall that $R_{\text{elem}} = R_H + \delta_i$ - the macroscopic segment is then governed by a linear approximation of this 1 GΩ network's internal transfer function below (which expands $\Delta R_N = \sum_{j=1}^{N} S_{N,j}\delta_j$, where $N$ refers to the total number of measurements, with our focus on the measurement example in the main text between North and East):

$$\Delta R_{\text{eq, North-East}} = \left(\frac{\partial R_{\text{eq}}}{\partial R_{\text{series}}}\right)\Delta R_{\text{series}} + \left(\frac{\partial R_{\text{eq}}}{\partial R_{\text{parallel}}}\right)\Delta R_{\text{parallel}} \tag{S2}$$

And recalling the star-mesh transformation for $R_{\text{eq}}$ yields a conventional Wye-Delta formula:

$$R_{\text{eq}} = R_{\text{series}} + R_{\text{series}} + \frac{(R_{\text{series}})^2}{R_{\text{parallel}}} \tag{S3}$$

Substituting the first-order summations for the series arm and the parallel shunt, the complete micro-to-macro mapping expands into this form:

$$\Delta R_{\text{eq, North-East}} = \left(2 + \frac{2R_{\text{series}}}{R_{\text{parallel}}}\right)\left(\sum_{j=1}^{40} \delta_{i,\text{series},j}\right) - \left(\frac{R_{\text{series}}^2}{R_{\text{parallel}}^2}\right)\left(\frac{1}{44^2}\sum_{k=1}^{44} \delta_{i,\text{parallel},k}\right) \tag{S4}$$

## 3. Determination of $\epsilon$

The observation window $\epsilon$ is expressed in ohms (Ω) and represents the standard deviation of a set of *N* measurements made. In the case of ideal measurements, $\epsilon$ should reflect the noise floor of the instrumentation. So, for instance, at 1 MΩ, assuming an ideal measurement and a relative standard deviation minimum of 100 nΩ/Ω (i.e., by use of a DSB), one obtains:

$$\epsilon_{1\,M\Omega} = (10^6\ \Omega) \times 10^{-7}\Omega/\Omega = 0.1\ \Omega$$

Within the framework of the SPQR criterion, $\epsilon$ abstractly defines the acceptance boundary in *N*-dimensional measurement space (where $n$ in this work is the number of unique partitions in a specific QHARS topology). Two things should be noted here: First, suppose $\epsilon$ (the spread of the measurements) is large. It then follows that evaluating the SPQR criterion yields a higher probability of artificiality, meaning that one's confidence that a QHARS device exhibits no breakdown of quantization intuitively decreases. Second, $\Delta R_N$ depends on $R_{\text{nominal}}$, which may or may not include errors that create offsets. Ultimately, the SPQR criterion gives weight to the deviation of the measurements ($\epsilon$) as opposed to offsets that may be common to all data (and thus accountable in an uncertainty budget).

## 4. Discussion on the Breakdown Parameter $\mu_n$

This section is largely reflective of the Supplemental Material of [1] and made accessible for this work due to direct relevance.

The loss of quantization is a relatively abrupt phase transition, and when the critical current is exceeded, mechanisms like inter-Landau-level scattering cause the system to rapidly heat and break down, making the longitudinal resistance jump suddenly by several orders of magnitude beyond the ohm range. This allows the value of 1 ohm to be a conservative representation of the loss of quantization ($\mu_i$), and therefore, we define $\mu_n$ to be the macroscopic signal that quantization has fundamentally collapsed somewhere along the active measurement path (presumably, and as a stringent bound, due to a single element failure).

The macroscopic breakdown parameter $\mu_n$ reflects the entire two-terminal measurement path and can be calculated using the chain rule to translate a microscopic elemental fault ($\mu_i \approx 1\ \Omega$) into a measurable shift in resistance. Let $R_i$ be the specific quantum Hall element experiencing this fault. Then $\mu_n$ is the total derivative of the measured equivalent resistance ($R_{\text{eq}}$) with respect to that specific faulty element:

$$\mu_n = \mu_i \left| \frac{dR_{\text{eq}}}{dR_i} \right|$$

$$\mu_n = \mu_i \left| \frac{\partial R_{\text{eq}}}{\partial R_{\text{partition}}} \times \frac{\partial R_{\text{partition}}}{\partial R_{\text{seg}}} \times \frac{\partial R_{\text{seg}}}{\partial R_i} \right| \tag{S5a}$$

To find $\mu_n$ systematically for any topology and layout, one expands the total derivative into three distinct topological layers within which the fault could exist: (1) highest level, (2) partition level, and (3) element level. More terms for additional layers of topological complexity may be added, but anything beyond what is in Eq. S5a is unlikely to be necessary.

The first term describes how the active measurement loop responds to a change in one of its partitions, the second term describes how a single (presumably faulty) partition's total resistance changes if one of its internal macroscopic segments (e.g., a specific series string or a parallel bank) experiences a breakdown. If the topology is simple (e.g., a partition of just a single string of elements in series), this derivative is simply 1. The third term translates the single faulty quantum Hall element into a shift of its immediate segment (if there are multiple segments in a partition). The use of this approach will be seen in more detail in the full QHARS calculations.

An important distinction must be made between the rotationally symmetric ($C_n$) and partition-resolved frameworks. The latter involves components that are highly different from one another compared with the initial investigations of the rotationally symmetric topologies. As such, one ambition of this work is to more accurately associate every observation window $\epsilon_k$ from a two-terminal measurement path $k$ with a conservative breakdown parameter $\mu_k$ of that two-terminal measurement path. In a rotationally symmetric topology, the measurement path retains a more profound level of symmetry than the partition-resolved case. To pursue this accurate association, careful bookkeeping is required to avoid counting Layer 1 ($\partial R_{\text{eq}}/\partial R_{\text{partition}}$) twice when transitioning from rotationally symmetric ($C_n$) topologies [1] to arbitrarily partitioned, subarray-resolved topologies:

1. *In the rotationally symmetric framework [1]:* Every two-terminal measurement was an identical symmetry rotation of the same full circuit topology. Consequently, the sensitivity matrix **S** functioned strictly as a binary topological incidence matrix (0s and 1s) indicating which branches were active in each measurement, while all three derivative layers of Eq. (S5a), including the star-mesh amplification $\partial R_{\mathrm{eq}}/\partial R_{\mathrm{partition}}$, were bundled into a single scalar breakdown parameter $\mu_n$ representing the entire two-terminal measurement path (selecting the minimum value across active subcomponents for mathematical conservatism).
2. *In the partition-resolved framework (this work):* A single QHARS device is characterized using heterogeneous two-terminal measurement configurations that combine ungrounded series paths (where $\partial R_{\mathrm{eq}}/\partial R_{\mathrm{partition}} = 1$) with grounded star-mesh paths (where $|\partial R_{\mathrm{eq}}/\partial R_{\mathrm{partition}}| \gg 1$). Because these circuit-level sensitivities vary from measurement to measurement and from partition to partition, Layer 1 above ($\partial R_{\mathrm{eq},k}/\partial R_{\mathrm{partition},j}$) is placed directly inside the sensitivity matrix **S** ($S_{k,j} = \partial R_{\mathrm{eq},k}/\partial R_{\mathrm{partition},j}$).
3. Consequently, to prevent double-counting the circuit-level star-mesh amplification between **S** and the Laplace prior $f(\vec{\delta})$, the partition breakdown parameter $\mu_j$ (denoted $\mu_n$ when identical across all $n$ partitions) is defined strictly by Layers 2 and 3 above, where the internal resistance shift of partition $j$ itself caused by a single-element breakdown:

$$\mu_j = \mu_i \left| \frac{\partial R_{\mathrm{partition},j}}{\partial R_{\mathrm{seg}}} \times \frac{\partial R_{\mathrm{seg}}}{\partial R_i} \right| \tag{S5b}$$

4. If a partition contains multiple inequivalent internal segments, the minimum value of Eq. (S5b) within that partition is selected for mathematical conservatism. For any partition composed purely of in-series elements or a single Hall element, both internal derivatives in Eq. (S5b) equal unity, giving $\mu_j = \mu_i = 1\ \Omega$ (denoted $\mu_a = 1\ \Omega$ for series arms and $\mu_s = 1\ \Omega$ for single-element shunts). For a non-series partition such as an $m$-element parallel shunt (see Design 2B, where $m = 8$), Eq. (S5b) yields $\mu_s = \mu_i/m^2 = 1/64\,\Omega$. When evaluating $P(A)$, the denominator product $\sqrt{\det(\boldsymbol{S}^T\boldsymbol{S})}\prod_{j=1}^{n}\mu_j$ combines Layer 1 (via $\sqrt{\det(\boldsymbol{S}^T\boldsymbol{S})}$) with Layers 2 and 3 (via $\prod_{j=1}^{n}\mu_j$), automatically pairing each measurement observation window $\epsilon_k$ with its full two-terminal breakdown scale while retaining the complete fault amplification of shunted star-mesh paths.

One thing to note here is that $\mu_n$ is intended to represent a physically meaningful quantity as opposed to a historical metrological compliance limit. For instance, one may call an element not fully quantized if it exhibits a longitudinal resistance in the mΩ range, but such a designation is rather arbitrary compared to a phase transition.

## 5. Example QHARS Calculation in Full

Suppose one evaluates the SPQR criterion for the first case in the main manuscript, namely, the 78-element series path and the two 40-element paths. The three measurements uniquely constrain the three macroscopic defect coordinates:

$$\boldsymbol{\delta} = \begin{pmatrix} \delta_1 \\ \delta_2 \\ \delta_3 \end{pmatrix}$$

where $\delta_1, \delta_2$ represent the two 39-element arms, and $\delta_3$ represents the single-element shunt. The three measurement deviations are

$$\begin{aligned} \Delta R_{78} &= \delta_1 + \delta_2 \\ \Delta R_{40,L} &= \delta_1 + \delta_3 \\ \Delta R_{40,R} &= \delta_2 + \delta_3 \end{aligned}$$

Thus,

$$\begin{pmatrix} \Delta R_{78} \\ \Delta R_{40,L} \\ \Delta R_{40,R} \end{pmatrix} = \begin{pmatrix} 1 & 1 & 0 \\ 1 & 0 & 1 \\ 0 & 1 & 1 \end{pmatrix} \begin{pmatrix} \delta_1 \\ \delta_2 \\ \delta_3 \end{pmatrix}$$

The determinant is $\det(S) = -2$, and therefore $\det(S^TS) = 4$:

$$\Phi \approx 1 - \frac{1}{\sqrt{\det(S^TS)}}\left(\frac{\epsilon_1\epsilon_2\epsilon_3}{\mu_n{}^3}\right) = 1 - \frac{1}{2}\frac{\epsilon_1\epsilon_2\epsilon_3}{\mu_n{}^3}$$

In the second case, where we keep the effective resistances of the QHARS device as:

$$R_1 = 39R_H + \delta_1$$
$$R_2 = 39R_H + \delta_2$$
$$R_3 = R_H + \delta_3$$

The defect vector remains the same as above and the nominal values are about 516.3 kΩ for each of the two measurements that include one arm and the single element shunt, and about 20.6 MΩ for the Wye-Delta configuration (as calculated using conventional formulas for the Wye-Delta transformation, or star-mesh transformation for more generalized cases). The measurement deviation and sensitivity matrix to first order are found using partial derivatives evaluated at the ideal point (recall that $R_\Delta = R_1 + R_2 + \frac{R_1 R_2}{R_3}$):

$$\Delta R_\Delta \approx \left.\frac{\partial R_\Delta}{\partial R_1}\right|_0 \delta_1 + \left.\frac{\partial R_\Delta}{\partial R_2}\right|_0 \delta_2 + \left.\frac{\partial R_\Delta}{\partial R_3}\right|_0 \delta_3$$

Since

$$\frac{\partial R_\Delta}{\partial R_1} = 1 + \frac{R_2}{R_3}$$
$$\frac{\partial R_\Delta}{\partial R_2} = 1 + \frac{R_1}{R_3}$$
$$\frac{\partial R_\Delta}{\partial R_3} = -\frac{R_1 R_2}{R_3^2}$$

one obtains the nominal sensitivity coefficients

$$\Delta R_\Delta \approx 40\delta_1 + 40\delta_2 - 1521\delta_3$$

Keeping $N = n$, then a sensitivity matrix $S'$ is:

$$S' = \begin{pmatrix} 40 & 40 & -1521 \\ 1 & 1 & 0 \\ 1 & 0 & 1 \end{pmatrix}$$

And $\det(S') = 1521$, where $S'$ is a minimal full-rank 3-measurement protocol. The small-$\epsilon$ approximation gives:

$$\Phi \approx 1 - \frac{\epsilon_\Delta \epsilon_{78} \epsilon_{40}}{1521\, \mu_a^2 \mu_s}$$

<u>On the 1.29 MΩ QHARS Design (101 elements)</u>:

Converting the empirical relative standard deviations in Fig. 3 into absolute observation windows $\epsilon = R_{\mathrm{nominal}} \times \sigma_{\mathrm{rel}}$ yields $\epsilon_{100} = (100R_H)(3564.122 \times 10^{-6}) \approx 4600\ \Omega$ for the 100-element series array [Fig. 3(a)] and $\epsilon_{51} = (51R_H)(8290.45817 \times 10^{-6}) \approx 5457\ \Omega$ for each of the two 51-element near-half subarrays [Fig. 3(b)]. The exact integration of Eqs. 5 and 6 must be employed. In the limit where $\epsilon_{100}, \epsilon_{51} \gg \mu_n$, all exponential decay terms $\exp(-\epsilon_k/\mu_n)$ vanish identically, causing the triple integral in Eq. 5 to saturate at unity: $P(A) \rightarrow 1$, corresponding to $\Phi \rightarrow 0.0$ %.

On the 1.01 MΩ QHARS Design (79 elements):

The relative standard deviations of the time-series measurements for the 40-element near-half series subarray [$R_{40} = 40R_H \approx 516.3$ kΩ, main text Fig. 4 (a)], the 78-element full series array [$R_{78} = 78R_H \approx 1.0067$ MΩ, Fig. 4 (b)], and the grounded-shunt Wye-Delta configuration [$R_\Delta = 1599R_H \approx 20.637$ MΩ, Fig. 4 (c)] are about 30.5 $\mu\Omega/\Omega$, 1.257 $\mu\Omega/\Omega$, and 10.7 $\mu\Omega/\Omega$, respectively. Multiplying each relative standard deviation by its corresponding nominal quantized resistance yields the experimental observation windows $\epsilon_{40} = 15.75\ \Omega$, $\epsilon_{78} = 1.27\ \Omega$, and $\epsilon_\Delta = 220.82\ \Omega$.

On the Comparison of the Subarray-Only vs. Star-Mesh Measurement Approach:

First, compute the subarray-only SPQR criterion using the DSB measurements of Figs. 4 (a) and 4 (b) alone (applying $\epsilon_{40} = 15.75\ \Omega$ twice for the two symmetric 40-element paths alongside $\epsilon_{78} = 1.27\ \Omega$). Because $\epsilon_{40}$ and $\epsilon_{78}$ exceed the unit series breakdown scale $\mu_n = 1\ \Omega$, Eq. 5 (and Eq. S9) is integrated directly with $\sqrt{\det(\mathbf{S}^T\mathbf{S})} = 2$, yielding $P(A)_{\text{sub}} = 0.5408$ and a modest confidence level of $\Phi_{\text{sub}} = 1 - P(A)_{\text{sub}} = 0.4592$ (45.92 %).

Next, compute the second SPQR criterion by incorporating the Y-Δ measurement of Fig. 4 (c) ($\epsilon_\Delta = 220.82\ \Omega$) together with Figs. 4 (a) and 4 (b). Grounding the central shunt amplifies the sensitivity matrix determinant from 2 to $|\det(\mathbf{S}')| = 1521$ for the minimal full-rank $3 \times 3$ measurement protocol, while the internal partition breakdown parameters remain $\mu_a = \mu_s = 1\ \Omega$ by Eq. (S5b). Because the $-1521\delta_3$ sensitivity entry compresses the effective shunt window to $\epsilon_\Delta/1521 = 0.145\ \Omega < \mu_s = 1\ \Omega$, scaling the exact integral by the determinant ratio $|\det(\mathbf{S}_{\text{sub}})|/|\det(\mathbf{S}')| = 2/1521$ suppresses $P(A)_{\text{Y-}\Delta}$ by nearly three orders of magnitude to $7.11 \times 10^{-4}$ ($\Phi_{\text{Y-}\Delta} = 99.929$ %), while direct 3D defect-space integration over $|\mathbf{S}'\vec{\delta}| \leq \vec{\epsilon}$ yields $P(A)_{\text{Y-}\Delta} = 0.0730$ ($\Phi_{\text{Y-}\Delta} = 92.70$ %) under a bilateral Laplace prior and $P(A)_{\text{Y-}\Delta} = 0.0551$ ($\Phi_{\text{Y-}\Delta} = 94.49$ %) under a nonnegative elemental prior. Moreover, under ideal bridge noise floors (1 $\text{n}\Omega/\Omega$ CCC for $R_{40}$, 100 $\text{n}\Omega/\Omega$ DSB for $R_{78}$, and $R_\Delta$, giving $\epsilon_{40} = 0.516\ \text{m}\Omega$, $\epsilon_{78} =$ $101\ \mathrm{m\Omega}$, and $\epsilon_\Delta = 2.06\ \Omega$), the small-window condition holds after shunt-sensitivity scaling and Eq. 4 yields $P(A)_{\text{Y-}\Delta} = 7.05 \times 10^{-8}$ ($\Phi_{\text{Y-}\Delta} = 99.999993$ %, exceeding $5\sigma$).

On the Four-Partition Multi-Decade Asymmetric Star-Mesh Array ($n = 4$):

To realize a single QHARS topology exhibiting $n = 4$ defect vectors that outputs near-decadic resistances across five consecutive orders of magnitude (from ~$10\ \mathrm{k\Omega}$ and ~$100\ \mathrm{k\Omega}$ up to ~$1\ \mathrm{M\Omega}$, ~$10\ \mathrm{M\Omega}$, and ~$100\ \mathrm{M\Omega}$), we formulate an asymmetric three-arm star-mesh network connected to a central node O with a single-element shunt to ground. The four partitions are:

$$R_1 = N_1 R_H + \delta_1 = 7R_H + \delta_1 \approx 90.345\ \mathrm{k\Omega} + \delta_1$$

(Arm 1: 7 elements in series, $\mu_1 = 1\ \Omega$)

$$R_2 = N_2 R_H + \delta_2 = 78R_H + \delta_2 \approx 1.00670\ \mathrm{M\Omega} + \delta_2$$

(Arm 2: 78 elements in series, $\mu_2 = 1\ \Omega$)

$$R_3 = N_3 R_H + \delta_3 = 96R_H + \delta_3 \approx 1.23901\ \mathrm{M\Omega} + \delta_3$$

(Arm 3: 96 elements in series, $\mu_3 = 1\ \Omega$)

$$R_4 = N_4 R_H + \delta_4 = 1R_H + \delta_4 \approx 12.906\ \mathrm{k\Omega} + \delta_4$$

(Central shunt: 1 Hall element, $\mu_4 = 1\ \Omega$)

Note that choosing $N_2 = 78$ keeps the highest Wye-Delta output $R_\Delta(2,3) = 7662\ R_H \approx 98.889\ \mathrm{M\Omega}$; choosing $N_2 = 79$ gives $R_\Delta(2,3) = 7759\ R_H \approx 100.141\ \mathrm{M\Omega}$, within 0.14% of $100\ \mathrm{M\Omega}$ in the DSB regime; and choosing $N_2 = 70$ gives $R_{1+2} = 77\ R_H \approx 993.79\ \mathrm{k\Omega}$ inside the $1\ \mathrm{M\Omega}$ CCC soft ceiling. All three variants are tabulated in Table S1.

Across its terminals, the 182-element (7 + 78 + 96 + 1) array provides five orders of magnitude of near-decade outputs:

1. $10^4\Omega$ decade: $R_4 = 1\ R_H \approx 12.906\ \mathrm{k\Omega}$ (CCC, $\epsilon_4 = 1.2906 \times 10^{-5}\Omega$)
2. $10^5\Omega$ decade (~$100\ \mathrm{k\Omega}$): $R_{1+4} = R_1 + R_4 = 8\ R_H \approx 103.251\ \mathrm{k\Omega}$ (3.25 % from $100\ \mathrm{k\Omega}$; CCC, $\epsilon_{1+4} = 1.0325 \times 10^{-4}\Omega$)

3. $10^6\Omega$ decade (~1 MΩ): $R_2$ = 78 $R_H$ ≈ 1.00670 MΩ, $R_{2+4}$ = 79 $R_H$ ≈ 1.01961 MΩ (DSB, $\epsilon_{2+4}$ = 1.0196 × $10^{-1}\Omega$)

4. $10^7\Omega$ decade (~10 MΩ): $R_\Delta(1,3)$ = $R_1$ + $R_3$ + $R_1$ $R_3$ / $R_4$ = 775 $R_H$ ≈ 10.00246 MΩ (0.025 % from 10 MΩ; DSB, $\epsilon_\Delta(1,3)$ = 1.0002 Ω)

5. $10^8\Omega$ decade (~100 MΩ): $R_\Delta(2,3)$ = $R_2$ + $R_3$ + $R_2$ $R_3$ / $R_4$ = 7662 $R_H$ ≈ 98.88887 MΩ (1.11 % from 100 MΩ; DSB, $\epsilon_\Delta(2,3)$ = 9.8889 Ω)

Additional auxiliary paths include $R_{1+2}$ = 85 $R_H$ ≈ 1.09704 MΩ (DSB, $\epsilon_{1+2}$= 1.0970 × $10^{-1}\Omega$), $R_{1+3}$ = 103 $R_H$ ≈ 1.32936 MΩ (DSB, $\epsilon_{1+3}$ = 1.3294 × $10^{-1}\Omega$), $R_{2+3}$ = 174 $R_H$ ≈ 2.24571 MΩ (DSB, $\epsilon_{2+3}$ = 2.2457 × $10^{-1}\Omega$), and $R_\Delta(1,2)$ = 631 $R_H$ ≈ 8.14394 MΩ (DSB, $\epsilon_\Delta(1,2)$ = 8.1439 × $10^{-1}\Omega$).

1. Subarray-Only Evaluation ($n$ = 4): Selecting four independent series-only measurements, specifically $R_{1+4}$ (~100 kΩ), $R_{2+4}$ (~1 MΩ), $R_{1+2}$ (~1.10 MΩ), and $R_{1+3}$ (~1.33 MΩ), where the first three form a 3-cycle on partitions {1, 2, 4} and the fourth links partition 3, yields the 4 × 4 sensitivity matrix:

$$S_{sub,4} = \begin{bmatrix} 1 & 0 & 0 & 1 \\ 0 & 1 & 0 & 1 \\ 1 & 1 & 0 & 0 \\ 1 & 0 & 1 & 0 \end{bmatrix}, |\det(S_{sub,4})| = 2, \sqrt{\det(S_{sub,4}^{\mathrm{T}} S_{sub,4})} = 2$$

$$P(A)_{\mathrm{sub},4} = \frac{\epsilon_{1+4}\epsilon_{2+4}\epsilon_{1+2}\epsilon_{1+3}}{2\mu_n^4} \approx 7.6765 \times 10^{-11}.$$

2. Star-Mesh (Y-Δ) Evaluation ($n$ = 4): For any pair of arms (a, b) measured with the shunt $R_4$ grounded, the first-order Taylor expansion of $R_\Delta(a,b)$ = $R_a$ + $R_b$ + $R_a$ $R_b$ / $R_4$ gives:

$$\Delta R_\Delta(2,3) \approx (1 + N_3)\delta_2 + (1 + N_2)\delta_3 - N_2 N_3 \delta_4 = 97\delta_2 + 79\delta_3 - 7488\delta_4$$

$$\Delta R_\Delta(1,3) \approx (1 + N_3)\delta_1 + (1 + N_1)\delta_3 - N_1 N_3 \delta_4 = 97\delta_1 + 8\delta_3 - 672\delta_4$$

$$\Delta R_\Delta(1,2) \approx (1 + N_2)\delta_1 + (1 + N_1)\delta_2 - N_1 N_2 \delta_4 = 79\delta_1 + 8\delta_2 - 546\delta_4$$

We evaluate three representative 4 × 4 Y-Δ measurement combinations:

- *Scheme 4-I* (Single Y-Δ + Three Series Paths: $R_{\Delta}(2,3)$, $R_{1+3}$, $R_{1+2}$, $R_{1+4}$): Replacing $\Delta R_2$+4 in $S_{sub,4}$ with $\Delta R_{\Delta}$(2,3) gives:

$$S'_{4,I} = \begin{bmatrix} 0 & 97 & 79 & -7488 \\ 1 & 0 & 1 & 0 \\ 1 & 1 & 0 & 0 \\ 1 & 0 & 0 & 1 \end{bmatrix}, |\det(S'_{4,I})| = 7312$$

$$P(A)_{Y\Delta,4,I} = (\epsilon_{\Delta}(2,3)\epsilon_{1+3}\epsilon_{1+2}\epsilon_{1+4})/(7312\mu_a^3\mu_s) \approx 2.0364 \times 10^{-12}$$

(37.7× lower $P(A)$ than subarray-only)

- *Scheme 4-II* (Two Decade Y-Δ Paths + Two Series Paths: $R_{\Delta}(2,3)$ [~100 MΩ], $R_{\Delta}(1,3)$ [~10 MΩ], $R_{2+3}$, $R_{1+4}$ [~100 kΩ]):

$$S'_{4,II} = \begin{bmatrix} 0 & 97 & 79 & -7488 \\ 97 & 0 & 8 & -672 \\ 0 & 1 & 1 & 0 \\ 1 & 0 & 0 & 1 \end{bmatrix}, |\det(S'_{4,II})| = 73{,}746$$

$$P(A)_{Y\Delta,4,II} = (\epsilon_{\Delta}(2,3)\epsilon_{\Delta}(1,3)\epsilon_{2+3}\epsilon_{1+4})/(73{,}746\mu_a^3\mu_s) \approx 3.1100 \times 10^{-12}$$

- *Scheme 4-III* (All Three DSB Y-Δ Paths and remaining CCC Series: $R_{\Delta}(2,3)$, $R_{\Delta}(1,3)$, $R_{\Delta}(1,2)$, $R_{1+4}$):

$$S'_{4,III} = \begin{bmatrix} 0 & 97 & 79 & -7488 \\ 97 & 0 & 8 & -672 \\ 79 & 8 & 0 & -546 \\ 1 & 0 & 0 & 1 \end{bmatrix}, |\det(S'_{4,III})| = 491{,}776$$

$$P(A)_{Y\Delta,4,III} = (\epsilon_{\Delta}(2,3)\epsilon_{\Delta}(1,3)\epsilon_{\Delta}(1,2)\epsilon_{1+4})/(491{,}776\mu_a^3\mu_s) \approx 1.6913 \times 10^{-12}$$

(45.4× lower $P(A)$ than subarray-only)

<u>On the Five-Partition QHARS Topologies ($n = 5$)</u>:

For $n = 5$ defect vectors, we formulate and analyze two complementary topologies that highlight distinct physical mechanisms of the subarray-resolved SPQR framework: (Design 2A) a Two-Stage Recursive Star-Mesh (Dual-Shunt Decadic Ladder, Fig. 1 (c)) network that goes up to and includes 1 GΩ, thus exercises at some point all three experimental apparatus regimes (CCC, direct current comparators, and DSB), and (Design 2B) a Subarray-Resolved Multi-Tier

Star-Mesh network with an 8-element parallel shunt (resolving the architecture of Fig. 1 (d) into five independent partitions) that explicitly demonstrates the multi-layer chain rule of Eq. S5b for non-series partitions ($\mu_s \neq 1\ \Omega$).

### A. *Two-Stage Recursive Star-Mesh (Dual-Shunt Decadic Ladder) Reaching* $1\ G\Omega$

In a two-stage recursive star-mesh ladder, input terminal $T_1$ connects via left series arm $R_1$ to internal node A (shunted to ground by $R_4$), node A connects via intermediate bridge arm $R_2$ to internal node B (shunted to ground by $R_5$), and node B connects via right series arm $R_3$ to output terminal $T_3$. Choosing an intermediate arm of $N_2$ = 9 elements with single-element shunts $N_4$ = $N_5$ = 1 creates an exact base-10 stage multiplier (1 + $R_2/R_4$ = 10). The five partitions are:

$$R_1 = 78R_H + \delta_1 \approx 1.00670\,\mathrm{M\Omega} + \delta_1, R_2 = 9R_H + \delta_2 \approx 116.158\,\mathrm{k\Omega} + \delta_2, R_3 = 88R_H + \delta_3 \approx 1.13576\,\mathrm{M\Omega} + \delta_3$$

$$R_4 = 1R_H + \delta_4 \approx 12.906\,\mathrm{k\Omega} + \delta_4$$

(Shunt 1 at Node A)

$$R_5 = 1\ R_H + \delta_5 \approx 12.906\,\mathrm{k\Omega} + \delta_5$$

(Shunt 2 at Node B)

Applying two successive star-mesh (Wye-Delta) transformations gives the exact two-stage effective resistance $R_{\Delta\Delta}$:

$$R_{\Delta\Delta} = R_1 + R_2 + R_3 + (R_1R_2)/R_4 + (R_1R_3)/R_4 + (R_1R_3)/R_5 + (R_2R_3)/R_5 + (R_1R_2R_3)/(R_4R_5)$$

This single 177-element ladder outputs near-exact decadic resistances across six consecutive orders of magnitude ($10^4\Omega$ to $10^9\Omega$), spanning all three experimental apparatus regimes:

1. $10^4\Omega$ decade: $R_4$ = $R_5$ = 1 $R_H$ ≈ $12.906\,\mathrm{k\Omega}$ (CCC, $\epsilon_4$ = 1.2906 × $10^{-5}\Omega$)
2. $10^5\Omega$ decade (~$100\,\mathrm{k\Omega}$): $R_{2+4}$ = $R_{2+5}$ = 10 $R_H$ ≈ $129.064\,\mathrm{k\Omega}$ (CCC, $\epsilon_{2+4}$ = 1.2906 × $10^{-4}\Omega$)

3. $10^6\Omega$ decade (~1 MΩ): $R_1$ = 78 $R_H \approx 1.00670\,\text{M}\Omega$, $R_{1+4}$ = 79 $R_H \approx 1.01961\,\text{M}\Omega$ (DSB, $\epsilon_{1+4}$ = 1.0196 × $10^{-1}\Omega$)

4. $10^7\Omega$ decade (~10 MΩ): $R_{\Delta 1}(1{,}2|R_4)$ = 789 $R_H \approx 10.18315\,\text{M}\Omega$ (DSB, $\epsilon_{\Delta 1}$ = 1.0183 Ω); $R_{\Delta 2}(2{,}3|R_5)$ = 889 $R_H \approx 11.47379\,\text{M}\Omega$ (DSB, $\epsilon_{\Delta 2}$ = 1.1474 Ω)

5. $10^8\Omega$ decade (~100 MΩ): $R_{\Delta,span}(R_4)$ = $R_1$ + ($R_2$+$R_3$) + $R_1(R_2 + R_3)/R_4$ = 7741 $R_H \approx$ $99.90847\,\text{M}\Omega$ (0.09 % from $100\,\text{M}\Omega$; DSB, $\epsilon_{\Delta,span4}$ = 9.9908 Ω)

6. $10^9\Omega$ decade (~1 GΩ): $R_{\Delta\Delta}(R_4, R_5)$ = 77,173 $R_H \approx 996.0259\,\text{M}\Omega \approx 0.9960\,\text{G}\Omega$ (0.40 % from 1 GΩ; DSB at $100\,\text{n}\Omega/\Omega$, $\epsilon_{\Delta\Delta}$ = 9.9603 × $10^1\Omega$)

(Note: As an alternative parameter set, choosing $N_1$ = 88, $N_2$ = 7, $N_3$ = 96, $N_4$ = $N_5$ = 1 yields $R_{2+4}$ = 8 $R_H \approx 103.251\,\text{k}\Omega$ [3.25% from $100\,\text{k}\Omega$], $R_{\Delta 2}$ = 775 $R_H \approx 10.00246\,\text{M}\Omega$ [0.025 % from $10\,\text{M}\Omega$], and $R_{\Delta\Delta}$ = 77,511 $R_H \approx 1.00039\,\text{G}\Omega$ [0.039 % from $1\,\text{G}\Omega$]; both parameter sets are included in Table S1.)

1. Subarray-Only Evaluation ($n = 5$): Using five ungrounded series measurements, $R_{1+2+3}$ = 175 $R_H \approx 2.2586\,\text{M}\Omega$ (DSB, $\epsilon_{123}$ = 2.2586 × $10^{-1}\Omega$), $R_{1+4}$ = 79 $R_H \approx 1.0196\,\text{M}\Omega$ (DSB, $\epsilon_{14}$ = 1.0196 × $10^{-1}\Omega$), $R_{2+4}$ = 10 $R_H \approx 129.06\,\text{k}\Omega$ (CCC, $\epsilon_{24}$ = 1.2906 × $10^{-4}\Omega$), $R_{2+5}$ = 10 $R_H \approx 129.06\,\text{k}\Omega$ (CCC, $\epsilon_{25}$ = 1.2906 × $10^{-4}\Omega$), and $R_{3+5}$ = 89 $R_H \approx$ $1.1487\,\text{M}\Omega$ (DSB, $\epsilon_{35}$ = 1.1487 × $10^{-1}\Omega$), yields:

$$S_{sub,5A} = \begin{bmatrix} 1 & 1 & 1 & 0 & 0 \\ 1 & 0 & 0 & 1 & 0 \\ 0 & 1 & 0 & 1 & 0 \\ 0 & 1 & 0 & 0 & 1 \\ 0 & 0 & 1 & 0 & 1 \end{bmatrix}, |\det(S_{sub,5A})| = 3$$

$$P(A)_{sub,5A} = (\epsilon_{123}\epsilon_{14}\epsilon_{24}\epsilon_{25}\epsilon_{35})/(3\mu_n^5) \approx 1.4688 \times 10^{-14}$$

2. Recursive Star-Mesh Evaluation ($n = 5$): Differentiating $R_{\Delta\Delta}$, $R_{\Delta,span}(R_4)$, and $R_{\Delta 2}(2{,}3|R_5)$ with respect to the five partition resistances at the nominal operating point ($N_1$ = 78, $N_2$ = 9, $N_3$ = 88, $N_4$ = $N_5$ = 1) gives the sensitivity rows:

$$\frac{\partial R_{\Delta\Delta}}{\partial \delta} = \begin{bmatrix} 1 + N_2 + 2N_3 + N_2N_3 \\ (N_1 + 1)(N_3 + 1) \\ 1 + 2N_1 + N_2 + N_1N_2 \\ -N_1(N_2 + N_3 + N_2N_3) \\ -N_3(N_1 + N_2 + N_1N_2) \end{bmatrix}^{\mathrm{T}}$$

At the nominal operating point, this row evaluates to $[978, 7031, 868, -69342, -69432]$.

$$\partial R_{\Delta,span}(R_4)/\,\partial\delta = [1 + N_2 + N_3 \quad 1 + N_1 \quad 1 + N_1 \quad -N_1(N_2 + N_3) \quad 0]$$
$$= [98 \quad 79 \quad 79 \quad -7566 \quad 0]$$

$$\partial R_{\Delta 2}(2,3|R_5)/\,\partial\delta = [0 \quad 1 + N_3 \quad 1 + N_2 \quad 0 \quad -N_2N_3] = [0 \quad 89 \quad 10 \quad 0 \quad -792]$$

Combining all five decadic outputs, namely $R_{\Delta\Delta}$ (~1 GΩ, DSB), $R_{\Delta,span}(R_4)$ (~100 MΩ, DSB), $R_{\Delta 2}$ (~10 MΩ, DSB), $R_{1+4}$ (~1 MΩ, DSB), and $R_{2+4}$ (~100 kΩ, CCC), into a 5 × 5 multi-decade sensitivity matrix $S'$ yields:

$$S'_{5A,dec} = \begin{bmatrix} 978 & 7031 & 868 & -69342 & -69432 \\ 98 & 79 & 79 & -7566 & 0 \\ 0 & 89 & 10 & 0 & -792 \\ 1 & 0 & 0 & 1 & 0 \\ 0 & 1 & 0 & 1 & 0 \end{bmatrix}$$

$$|\det(S'_{5A,dec})| = 4{,}404{,}668{,}928 \approx 4.4047 \times 10^9$$

$$P(A)_{Y\Delta,5A,dec} = (\epsilon_{\Delta\Delta}\epsilon_{\Delta,span4}\epsilon_{\Delta 2}\epsilon_{14}\epsilon_{24})/(4.4047 \times 10^9 \mu_n^5) \approx 3.4112 \times 10^{-15}$$

(4.31× lower $P(A)$ than subarray-only)

Furthermore, if one replaces the 1 GΩ DSB measurement (which carries a $100\,\mathrm{n\Omega/\Omega}$ noise floor) with the full series path $R_{1+2+3}$ while retaining the two DSB star-mesh measurements $R_{\Delta,span}(R_4)$ (~100 MΩ) and $R_{\Delta 2}$ (~10 MΩ) along with $R_{1+4}$ (~1 MΩ) and $R_{2+5}$ (~100 kΩ), then $|\det(S'_{5A,\mathrm{DSB}})|$ = 6,606,535 and $P(A)$ drops to $5.1572 \times 10^{-16}$ (a 28.5-fold improvement over the subarray-only configuration).

*B. Multi-Tier Star-Mesh Array*

As a second $n = 5$ case that directly illustrates the multi-layer chain rule of Eq. S5b for a non-series partition, we resolve the multi-tier star-mesh topology of Fig. 1 (d) into five independent subarray partitions meeting at central node O: two 3-element lower-tier series arms ($R_1$ = 3 $R_H$ + $\delta_1$, $R_2$ = 3 $R_H$ + $\delta_2$ $\approx$ 38.719 kΩ), two 31-element upper-tier series arms ($R_3$ = 31 $R_H$ + $\delta_3$, $R_4$ = 31 $R_H$ + $\delta_4$ $\approx$ 400.099 kΩ), and a central shunt partition consisting of $k = 8$ Hall elements in parallel ($R_5$ = $R_H$/8 + $\delta_5$ $\approx$ 1.6133 kΩ; total 76 elements). By Eq. S5b, while the four series arms have $\mu_a = \mu_1 = \mu_2 = \mu_3 = \mu_4 = 1\Omega$, a single-element breakdown ($\mu_i$ = 1Ω) within the 8-element parallel shunt yields a partition-level breakdown parameter:

$$\mu_s = \mu_5 = \mu_i \left|\frac{\partial R_5}{\partial R_i}\right|$$

$$\mu_s = (1\,\Omega) \times \left|\frac{\partial}{\partial R_i}(R_i^{-1} + 7R_H^{-1})^{-1}\right|_0$$

$$\mu_s = \frac{1\,\Omega}{8^2} = \frac{1}{64}\,\Omega = 0.015625\,\Omega.$$

Thus, the product of the five partition breakdown parameters is $\mu_a^4\mu_s = 1/64\Omega^5 = 0.015625\,\Omega^5$. Across its five branches, this 76-element topology outputs five orders of magnitude of resistance: $R_5$ = $R_H$/8 $\approx$ 1.613 kΩ (CCC), $R_{1+5}$ = $R_{2+5}$ = 3.125 $R_H$ $\approx$ 40.333 kΩ (CCC, $\epsilon_{15}$ = 4.0333 × $10^{-5}$Ω), $R_{1+2}$ = 6 $R_H$ $\approx$ 77.438 kΩ (CCC, $\epsilon_{12}$ = 7.7438 × $10^{-5}$Ω), $R_{1+3}$ = 34 $R_H$ $\approx$ 438.818 kΩ (CCC, $\epsilon_{13}$ = 4.3882 × $10^{-4}$Ω), $R_{3+4}$ = 62 $R_H$ $\approx$ 800.197 kΩ (CCC, $\epsilon_{34}$ = 8.0020 × $10^{-4}$Ω), plus three decadic Wye-Delta outputs when $R_5$ = $R_H$/8 is grounded:

1. Lower-tier Y-Δ (3 vs. 3): $R_\Delta(1,2)$ = (3 + 3 + 8×3×3) $R_H$ = 78 $R_H$ $\approx$ 1.00670 MΩ (0.67 % from 1 MΩ; DSB, $\epsilon_\Delta(1,2)$ = 1.0067 × $10^{-1}$Ω)

2. Cross-tier Y-Δ (3 vs. 31): $R_\Delta(1,3)$ = (3 + 31 + 8×3×31) $R_H$ = 778 $R_H$ $\approx$ 10.04118 MΩ (0.41 % from 10 MΩ; DSB, $\epsilon_\Delta(1,3)$ = 1.0041 Ω)

3. Upper-tier Y-Δ (31 vs. 31): $R_\Delta(3,4)$ = (31 + 31 + 8×31×31) $R_H$ = 7750 $R_H$ $\approx$ 100.02463 MΩ (0.025 % from 100 MΩ; DSB, $\epsilon_\Delta(3,4)$ = 1.0002 × $10^{1}$Ω)

4. Subarray-Only Evaluation (Design 2B): Note that for a star topology, choosing four series measurements among the four arms {1, 2, 3, 4} (such as $R_{1+2}$, $R_{3+4}$, $R_{1+3}$, $R_{2+4}$) forms an even 4-cycle whose incidence rows are linearly dependent. A full-rank 5 × 5 subarray sensitivity matrix requires an odd cycle (such as the 3-cycle $\{R_{1+2}, R_{1+5}, R_{2+5}\}$ combined with $R_{1+3}$ and $R_{3+4}$), giving $|\det(S_{sub,5B})| = 2$ and:

$$P(A)_{\text{sub},5B} = \frac{\epsilon_{12}\epsilon_{34}\epsilon_{13}\epsilon_{15}\epsilon_{25}}{2\mu_a^4\mu_s}$$

$$P(A)_{\text{sub},5B} \approx 1.4155 \times 10^{-18}$$

5. Star-Mesh (Y-Δ) Evaluation (Design 2B): Because $\partial R_\Delta(a,b)/\,\partial R_5 = -R_aR_b/R_5^2 = -64N_aN_b$, the sensitivity to $\delta_5$ is amplified, which exactly cancels the $\mu_s$ = 1/64 Ω factor. Using the three decadic Y-Δ measurements $R_\Delta(3,4)$ (~100 MΩ), $R_\Delta(1,3)$ (~10 MΩ), and $R_\Delta(1,2)$ (~1 MΩ) together with $R_{2+5}$ (40.33 kΩ) and $R_{4+5}$ (401.71 kΩ, CCC, $\epsilon_{45}$ = 4.0171 × $10^{-4}$Ω) yields:

$$S'_{5B,Y\Delta} = \begin{bmatrix} 0 & 0 & 249 & 249 & -61504 \\ 249 & 0 & 25 & 0 & -5952 \\ 25 & 25 & 0 & 0 & -576 \\ 0 & 1 & 0 & 0 & 1 \\ 0 & 0 & 0 & 1 & 1 \end{bmatrix}$$

$$|\det(S'_{5B,Y\Delta})| = 38{,}807{,}026, |\det(S'_{5B,Y\Delta})|\mu_a^4\mu_s = 38{,}807{,}026/64 \approx 606{,}359.78$$

$$P(A)_{Y\Delta,5B} = (\epsilon_\Delta(3{,}4)\epsilon_\Delta(1{,}3)\epsilon_\Delta(1{,}2)\epsilon_{25}\epsilon_{45})/(|\det(S'_{5B,Y\Delta})|\mu_a^4\mu_s) \approx 2.7052 \times 10^{-16}$$

If instead the single DSB cross-tier measurement $R_\Delta(1{,}3)$ = 10.041 MΩ ($\epsilon_\Delta(1{,}3)$ = 1.0041 Ω) is paired with the four CCC series measurements $R_{1+2}$, $R_{1+3}$, $R_{2+3}$, and $R_{3+4}$, then $|\det(S')|$ = 11,904 and $P(A)$ = 6.442 × $10^{-17}$. Table S1 summarizes the complete parameters and SPQR results across all evaluated topologies with $n$ = 4 or 5.

**Table S1. Summary of Subarray-Resolved SPQR Parameters and Ideal-Apparatus Evaluations**

| QHARS Topology (Elements) | Measurement Protocol and Apparatus Regimes | $n$ ($N$) | $\sqrt{\det(S^T S)}$ | $P(A)$ |
|---|---|---|---|---|
| Design 1 (7+78+96+1) | Subarray Series (1 CCC, 3 DSB) | 4 (4) | 2 | $7.677 \times 10^{-8}$ |
| Design 1 (7+78+96+1) | Y-Δ Scheme I: 1 Y-Δ + 3 Ser (1 CCC, 3 DSB) | 4 (4) | 7,312 | $2.036 \times 10^{-9}$ |
| Design 1 (7+78+96+1) | Y-Δ Scheme II: 2 Y-Δ + 2 Ser (1 CCC, 3 DSB) | 4 (4) | 73,746 | $3.110 \times 10^{-9}$ |
| Design 1 (7+78+96+1) | Y-Δ Scheme III: 3 Y-Δ + 1 Ser (1 CCC, 3 DSB) | 4 (4) | 491,776 | $1.691 \times 10^{-9}$ |
| Design 1 Variant (7+79+96+1, 100.14 MΩ) | Y-Δ Scheme III: 3 Y-Δ + 1 Ser (1 CCC, 3 DSB) | 4 (4) | 497,992 | $1.713 \times 10^{-9}$ |
| Design 1 Variant (7+70+96+1, 993.8 kΩ) | Y-Δ Scheme I: 1 Y-Δ + 3 Ser (2 CCC, 2 DSB) | 4 (4) | 6,552 | $1.850 \times 10^{-11}$ |
| Design 2A: 2-Stage Ladder (78+9+88+1+1) | Subarray Series (2 CCC, 3 DSB) | 5 (5) | 3 | $1.469 \times 10^{-11}$ |
| Design 2A: 2-Stage Ladder (78+9+88+1+1) | Star-Mesh 5-Decade (1 CCC, 4 DSB) | 5 (5) | $4.405 \times 10^{9}$ | $3.411 \times 10^{-12}$ |
| Design 2A: 2-Stage Ladder (78+9+88+1+1) | Star-Mesh DSB-Opt (1 CCC, 4 DSB) | 5 (5) | $6.607 \times 10^{6}$ | $5.157 \times 10^{-12}$ |
| Design 2A Variant (88+7+96+1+1, 1.0004 GΩ) | Star-Mesh Scheme B (2 CCC, 3 DSB) | 5 (5) | $1.209 \times 10^{6}$ | $2.175 \times 10^{-13}$ |
| Design 2B: Multi-Tier (3+3+31+31+1/8) | Subarray Series (5 CCC) | 5 (5) | 2 | $1.416 \times 10^{-18}$ |
| Design 2B: Multi-Tier (3+3+31+31+1/8) | Y-Δ 10 MΩ DSB + 4 CCC Series | 5 (5) | 11,904 | $6.442 \times 10^{-17}$ |
| Design 2B: Multi-Tier (3+3+31+31+1/8) | Y-Δ 3-Decade (2 CCC, 3 DSB) | 5 (5) | $3.881 \times 10^{7}$ | $2.705 \times 10^{-14}$ |

## 6. Integration Assessment Without Assuming $\epsilon \ll \mu_n$

Starting with the generalized probability of artificiality $P(A)$:

$$P(A) = \frac{1}{2^n\sqrt{\det(\mathbf{S}^T\mathbf{S})}\prod_{j=1}^{n}\mu_j}\int_{\mathcal{P}} \exp\left(-\sum_{j=1}^{n}\frac{\left|\left(\mathbf{S}^{-1}\Delta\vec{R}\right)_j\right|}{\mu_j}\right) d^n(\Delta R)$$

(S6)

This integral becomes less straightforward if one does not assume that $\epsilon_k \ll \mu_j$. For the three-partition ($n = 3$) topology, the joint probability density function in defect space is:

$$f(\vec{\delta}) = \frac{1}{8\mu_1\mu_2\mu_3}\exp\left(-\frac{|\delta_1|}{\mu_1} - \frac{|\delta_2|}{\mu_2} - \frac{|\delta_3|}{\mu_3}\right)$$

(S7)

Let the three observation windows be $\epsilon_1, \epsilon_2, \epsilon_3$ corresponding to the three two-terminal subarray measurements $\Delta R_1 = \delta_1 + \delta_2$, $\Delta R_2 = \delta_1 + \delta_3$, and $\Delta R_3 = \delta_2 + \delta_3$, with sensitivity matrix $S_{\text{sub}}$ ($\sqrt{\det(\mathbf{S}^T\mathbf{S})} = |\det(S_{\text{sub}})| = 2$). Using the exact inverse matrix

$$S_{\text{sub}}^{-1} = \frac{1}{2}\begin{pmatrix} 1 & 1 & -1 \\ 1 & -1 & 1 \\ -1 & 1 & 1 \end{pmatrix}$$

and changing variables from defect space ($\Omega_\delta = \{\vec{\delta}: |\Delta R_k| \le \epsilon_k,\ k = 1,2,3\}$) to measurement-deviation space via $d^3\delta = \left[\sqrt{\det(\mathbf{S}^T\mathbf{S})}\right]^{-1} d\Delta R_1\, d\Delta R_2\, d\Delta R_3$ gives:

$$P(A) = \frac{1}{8\sqrt{\det(\mathbf{S}^T\mathbf{S})}\ \mu_1\mu_2\mu_3}\int_{-\epsilon_1}^{\epsilon_1}\int_{-\epsilon_2}^{\epsilon_2}\int_{-\epsilon_3}^{\epsilon_3} \exp\left[-\frac{|\Delta R_1 + \Delta R_2 - \Delta R_3|}{2\mu_1} - \frac{|\Delta R_1 - \Delta R_2 + \Delta R_3|}{2\mu_2} - \frac{|-\Delta R_1 + \Delta R_2 + \Delta R_3|}{2\mu_3}\right] d\Delta R_3\, d\Delta R_2\, d\Delta R_1$$

(S8)

For subarray-only (series) measurements, a shared breakdown parameter holds: $\mu_1 = \mu_2 = \mu_3 = \mu_n$. Because the linear transformation $\Delta\vec{R} = S_{\text{sub}}\vec{\delta}$ maps $\Omega_\delta$ one-to-one onto $[-\epsilon_1, \epsilon_1] \times [-\epsilon_2, \epsilon_2] \times [-\epsilon_3, \epsilon_3]$, the inner integral is most transparently evaluated by integrating first over $\delta_3$ at fixed $(\delta_1, \delta_2)$ satisfying $|\delta_1 + \delta_2| \le \epsilon_1$. The remaining two measurement bounds $|\delta_1 + \delta_3| \le \epsilon_2$ and $|\delta_2 + \delta_3| \le \epsilon_3$ restrict $\delta_3$ to the interval $L(\delta_1, \delta_2) \le \delta_3 \le U(\delta_1, \delta_2)$, where:

$$L(\delta_1, \delta_2) = \max(-\epsilon_2 - \delta_1, -\epsilon_3 - \delta_2), \qquad U(\delta_1, \delta_2) = \min(\epsilon_2 - \delta_1, \epsilon_3 - \delta_2)$$

Whenever $L(\delta_1, \delta_2) < U(\delta_1, \delta_2)$, the innermost integral

$$J_3(\delta_1, \delta_2) = \int_{L(\delta_1,\delta_2)}^{U(\delta_1,\delta_2)} \frac{1}{2\mu_n} \exp\left(-\frac{|\delta_3|}{\mu_n}\right) d\delta_3$$

splits at $\delta_3 = 0$ into three elementary cases:

$$J_3(\delta_1, \delta_2) = \begin{cases} 1 - \frac{1}{2} e^{L/\mu_n} - \frac{1}{2} e^{-U/\mu_n}, & L \le 0 \le U \\ \frac{1}{2}\left(e^{U/\mu_n} - e^{L/\mu_n}\right), & L < U < 0 \\ \frac{1}{2}\left(e^{-L/\mu_n} - e^{-U/\mu_n}\right), & 0 < L < U \end{cases}$$

Integrating $J_3(\delta_1, \delta_2)$ over the remaining coordinates $(\delta_1, \delta_2)$ with $|\delta_1 + \delta_2| \le \epsilon_1$ involves only elementary exponential integrals $\int_a^b e^{\lambda y}\, dy$. For a symmetric three-partition QHARS device where $\epsilon_1$ is the full-series observation window ($\epsilon_{100}$ or $\epsilon_{78}$) and $\epsilon_2 = \epsilon_3$ is the arm-plus-shunt observation window ($\epsilon_{51}$ or $\epsilon_{40}$) with $\epsilon_1 \le \epsilon_2$, this procedure yields the exact closed-form expression:

$$P(A) = 1 - \left(1 + \frac{\epsilon_1}{2\mu_n}\right) e^{-\frac{\epsilon_1}{\mu_n}} + \left(\frac{1}{3} - \frac{\epsilon_1}{\mu_n}\right) e^{-\frac{\epsilon_2}{\mu_n}} - e^{-\frac{\epsilon_1 + 2\epsilon_2}{2\mu_n}} + e^{-\frac{\epsilon_1 + \epsilon_2}{\mu_n}} - \frac{1}{3} e^{-\frac{3\epsilon_1 + 2\epsilon_2}{2\mu_n}} + \frac{1}{2} e^{-\frac{2\epsilon_2}{\mu_n}} - \frac{2}{3} e^{-\frac{\epsilon_1 + 4\epsilon_2}{2\mu_n}} + \frac{1}{6} e^{\frac{\epsilon_1 - 2\epsilon_2}{\mu_n}} \tag{S9}$$

When $\epsilon_2 \gg \mu_n$ while $\epsilon_1 \sim \mu_n$ (as in the 79-element DSB subarray measurements where $\epsilon_{78} = 1.27\ \Omega$, $\epsilon_{40} = 15.75\ \Omega$, and $\mu_n = 1\ \Omega$), all terms proportional to $e^{-\epsilon_2/\mu_n}$ and $e^{-2\epsilon_2/\mu_n}$ are negligible ($\sim 10^{-7}$), and Eq. (S9) simplifies to $P(A) \approx 1 - \left(1 + \frac{\epsilon_1}{2\mu_n}\right) e^{-\epsilon_1/\mu_n} = 0.5408$ ($\Phi_{\text{sub}} = 45.92\ \%$).

Furthermore, when all three observation windows are equal ($\epsilon_1 = \epsilon_2 = \epsilon_3 = \epsilon$), Eq. (S9) reduces to the closed-form exact result:

$$P(A) = 1 - \frac{1 + 3\left(\frac{\epsilon}{\mu_n}\right)}{2} e^{-\left(\frac{\epsilon}{\mu_n}\right)} - e^{-\frac{3}{2}\left(\frac{\epsilon}{\mu_n}\right)} + \frac{3}{2} e^{-2\left(\frac{\epsilon}{\mu_n}\right)} - e^{-\frac{5}{2}\left(\frac{\epsilon}{\mu_n}\right)} \tag{S10}$$

Expanding Eq. (S9) or Eq. (S10) in the small-window limit $\left(\frac{\epsilon_k}{\mu_n}\right) \ll 1$ gives:

$$P(A) = \frac{1}{\sqrt{\det(\mathbf{S}^T\mathbf{S})}}\left(\frac{\epsilon_1\epsilon_2\epsilon_3}{\mu_n^3}\right) + O\left[\left(\frac{\epsilon}{\mu_n}\right)^4\right]$$

recovering Eq. (3) of the main text.

## 7. Proposed QHARS Designs

From the main text, Figure 1 (b)-(d) are laid out as lithographic templates.

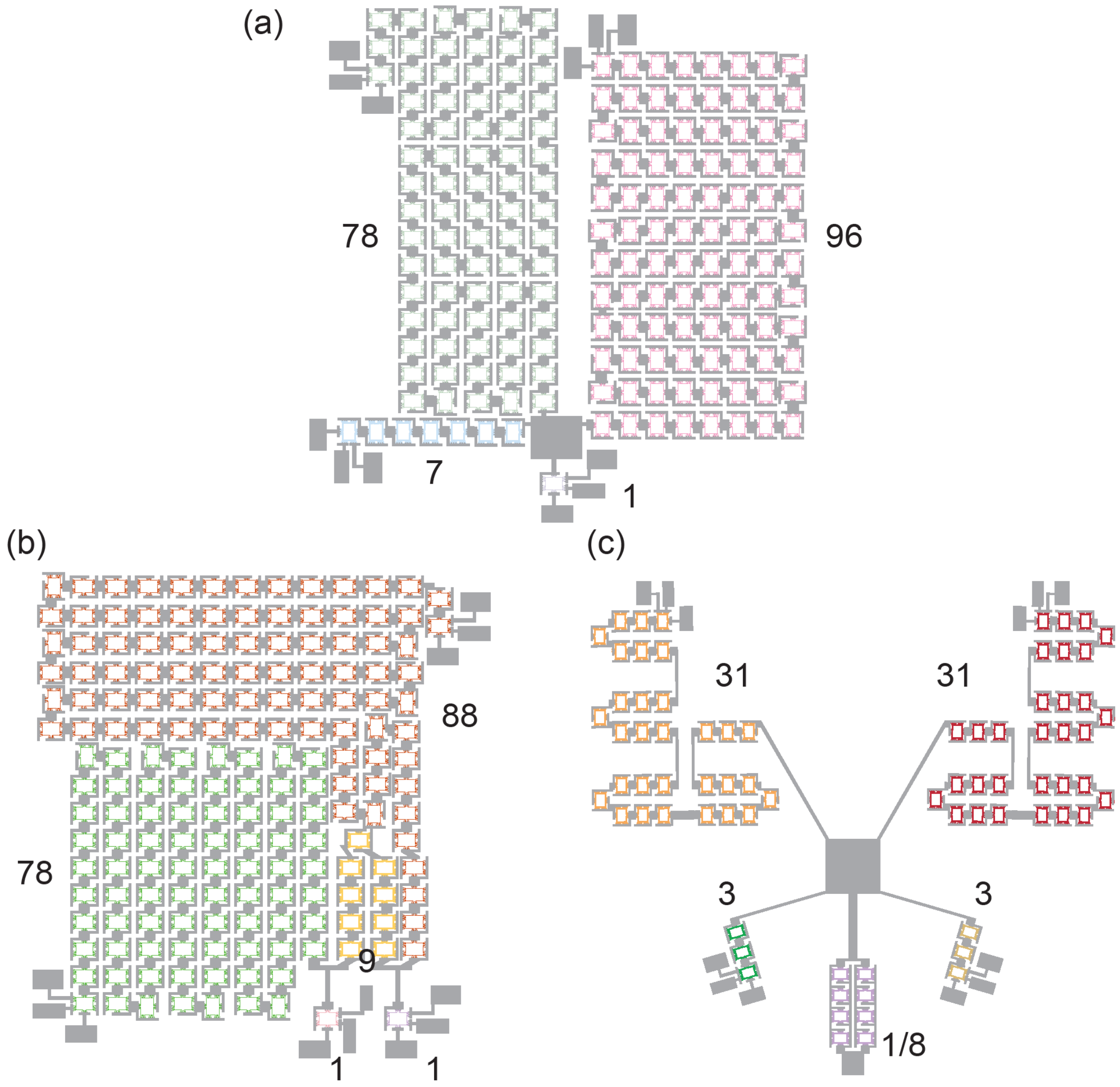

Figure 1-SM. Lithographic layout for the design of main text Figure 1 (b) to (d).